\documentclass[aps,prx,twocolumn,floatfix,superscriptaddress]{revtex4-2}

\usepackage{amssymb,amsmath,amstext}                
\usepackage{graphicx}
\usepackage{epstopdf}                                               
\usepackage{color}                                                     
\usepackage{bm}                                                        
\usepackage{appendix}

\usepackage[utf8]{inputenc}
\usepackage{lipsum}

\usepackage{tikz,xcolor}

\usepackage{ulem}
\usepackage{latexsym}
\usepackage[colorlinks=true,citecolor=blue,linkcolor=magenta]{hyperref}
\usepackage{cleveref}
\usepackage{soul}
\usepackage{amsmath}
\usepackage{float}

\newcommand{\vect}[1]{\mathbf{#1}}

\def\be{\begin{equation}}
\def\ee{\end{equation}}
\def\bea{\begin{eqnarray}}
\def\eea{\end{eqnarray}} 
\def\ra{\rangle}

\def\bi{\begin{itemize}}
\def\ei{\end{itemize}}

\definecolor{dgreen} {RGB}{78,138,21}

\definecolor{giergiel} {RGB}{0,128,88}

\definecolor{purple} {RGB}{128,0,160}

\usepackage{orcidlink}

\begin{document}
\title{Time-tronics: from temporal printed circuit board to quantum computer}
\author{Krzysztof Giergiel\,\orcidlink{0000-0003-3297-796X}}
\affiliation{Instytut Fizyki Teoretycznej, Uniwersytet Jagiello\'{n}ski, 
ulica Profesora Stanislawa Lojasiewicza 11, PL-30-348 Krak{ó}w, Poland}
\affiliation{Optical Sciences Centre, Swinburne University of Technology, Melbourne
Victoria 3122, Australia}
\affiliation{CSIRO, Manufacturing, Research Way, Clayton, Victoria 3168, Australia}
\author{Peter Hannaford\,\orcidlink{0000-0001-9896-7284}}
\affiliation{Optical Sciences Centre, Swinburne University of Technology, Melbourne, Victoria 
3122, Australia}
\author{Krzysztof Sacha\,\orcidlink{0000-0001-6463-0659}}
\thanks{krzysztof.sacha@uj.edu.pl}
\affiliation{Instytut Fizyki Teoretycznej, Uniwersytet Jagiello\'{n}ski, 
ulica Profesora Stanislawa Lojasiewicza 11, PL-30-348 Krak{ó}w, Poland}

\begin{abstract}
Time crystalline structures can be created in periodically driven systems. They are temporal lattices which can reveal different condensed matter behaviors ranging from Anderson localization in time to temporal analogues of many-body localization or topological insulators. However, the potential practical applications of time crystalline structures have yet to be explored. Here, we pave the way for time-tronics where temporal lattices are like printed circuit boards for realization of a broad range of quantum devices. The elements of these devices can correspond to structures of dimensions higher than three and can be arbitrarily connected and reconfigured at any moment. Moreover, our approach allows for the construction of a quantum computer, enabling quantum gate operations for all possible pairs of qubits. Our findings indicate that the limitations faced in building devices using conventional spatial crystals can be overcome by adopting crystalline structures in time.
\end{abstract}

\date{\today}

\maketitle

\section{Introduction}

Time crystals, much like their spatial counterparts, spontaneously form through the self-organization of many-body systems but in the time domain \cite{Wilczek2012,Bruno2013b,Watanabe2015,Kozin2019}. In periodically driven systems, discrete time crystals can emerge, evolving spontaneously with a period longer than  that dictated by the periodic perturbation \cite{Sacha2015,Khemani16,ElseFTC,Zhang2017,
Choi2017,Pal2018,Rovny2018,
Smits2018,Mi2022,Randall2021,Frey2022,
Kessler2020,Kyprianidis2021,
Xu2021,Taheri2020,Bao2024,Kazuya2024,Liu2024,Liu2024a}. In periodically disturbed systems, it is also possible to create time lattices, not resulting from self-organization, but from the application of appropriate temporal resonant perturbations \cite{Guo2013,Sacha15a,Buchleitner2002}. Studies have shown that such time crystalline structures can manifest a wide range of phases known in the physics of condensed matter, such as Anderson localization and many-body localization in time, topological insulators in time, and Mott insulators in the time domain \cite{SachaTC2020,GuoBook2021,Hannaford2022}.

From the inception of the field of time crystals, questions have arisen regarding how crystalline structures in time can be practically utilized and whether they carry greater potential for practical applications than conventional spatial crystals \cite{SachaTC2020,Hannaford2022a,Zaletel2023,
Bomantara2018,Estarellas2020,
Carollo2020,Lyu2020,
Iemini2023,montenegro2023,Bao2024,
Yousefjani2024}. In this article, we demonstrate that it is possible to achieve time-tronics, where time crystalline structures serve as printed circuit boards as in electronics, allowing the design and realization of a broad range of quantum devices.

We first illustrate how resonantly driven ultra-cold atoms can create a time crystalline structure where the connection of arbitrary sites through the tunneling of atoms between sites or atom-atom interactions can be realized (Sec.~\ref{tpcb}). As any connections between sites can be controlled, it is possible to realize a broad class of quantum devices, ranging from one-, two-, three- or higher-dimensional structures to more exotic objects like a Klein bottle, all of which can be connected in an arbitrary way, and the entire system can be reconfigured at any time during an experiment.
Next, in Sec.~\ref{qcsec}, we focus on a specific example, a quantum computer \cite{Preskill2018}. We demonstrate that a temporal printed circuit board can host qubits, where all single-qubit operations can be realized and a controlled-Z gate can be performed between all possible qubit pairs, meeting the conditions for a universal quantum computer. Section~\ref{sandc} contains the summary and conclusions.

\section{Temporal printed circuit board}
\label{tpcb}

In this section, we describe how resonantly driven ultracold atoms can be used to construct a temporal printed circuit board --- a device based on a temporal lattice, in which atomic tunneling between arbitrary lattice sites can be implemented, interactions between atoms occupying any chosen pairs of sites can be controlled, and the entire system can be dynamically reconfigured during the course of the experiment. We begin by introducing the concept of a temporal crystalline structure, then explain how tunneling and interactions can be controlled, and finally discuss the idea of the circuit board.

\subsection{Time crystalline structure}

\begin{figure*}[t]
\includegraphics[width=0.329\textwidth]{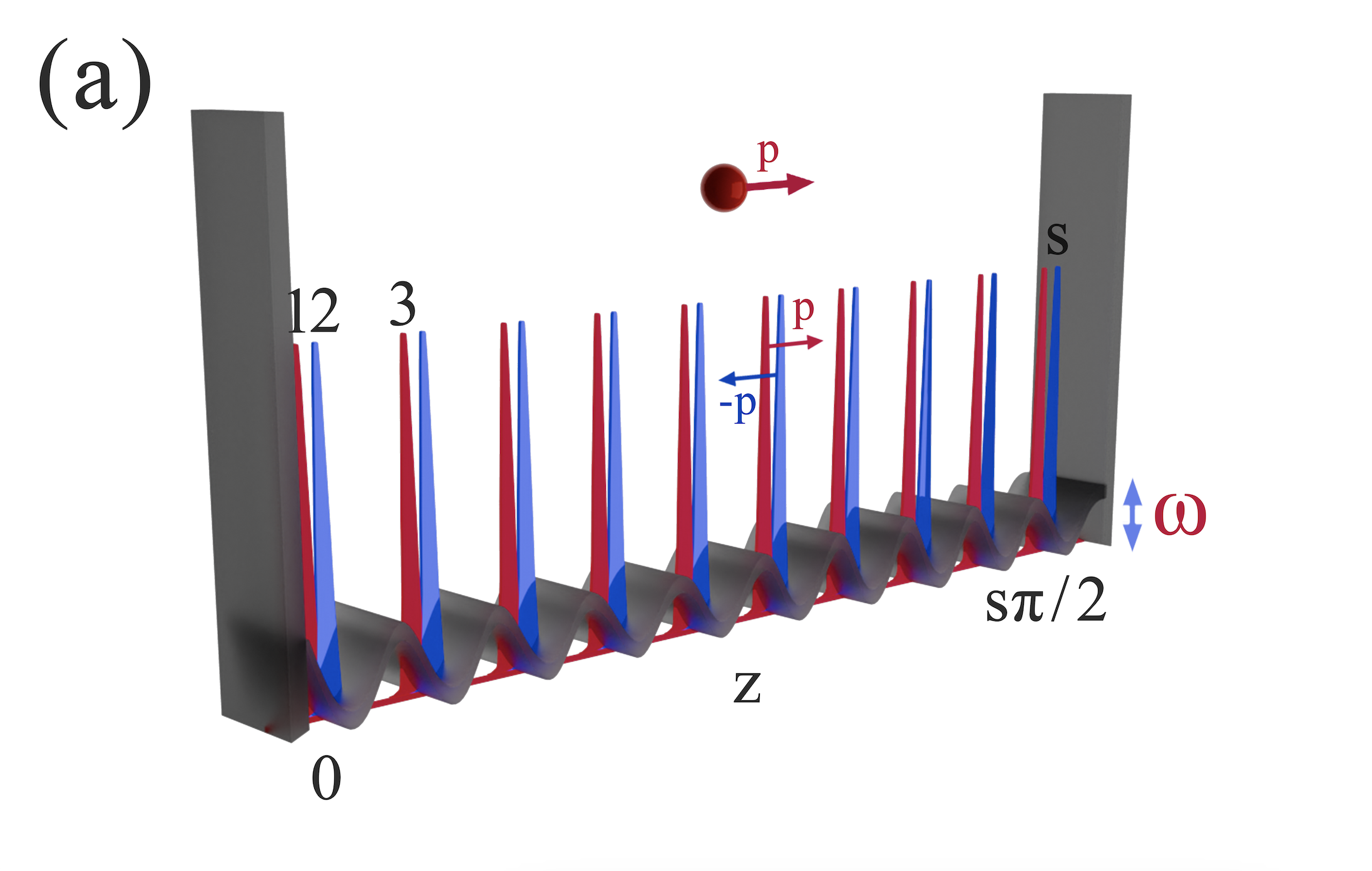}
\hfill
\includegraphics[width=0.329\textwidth]{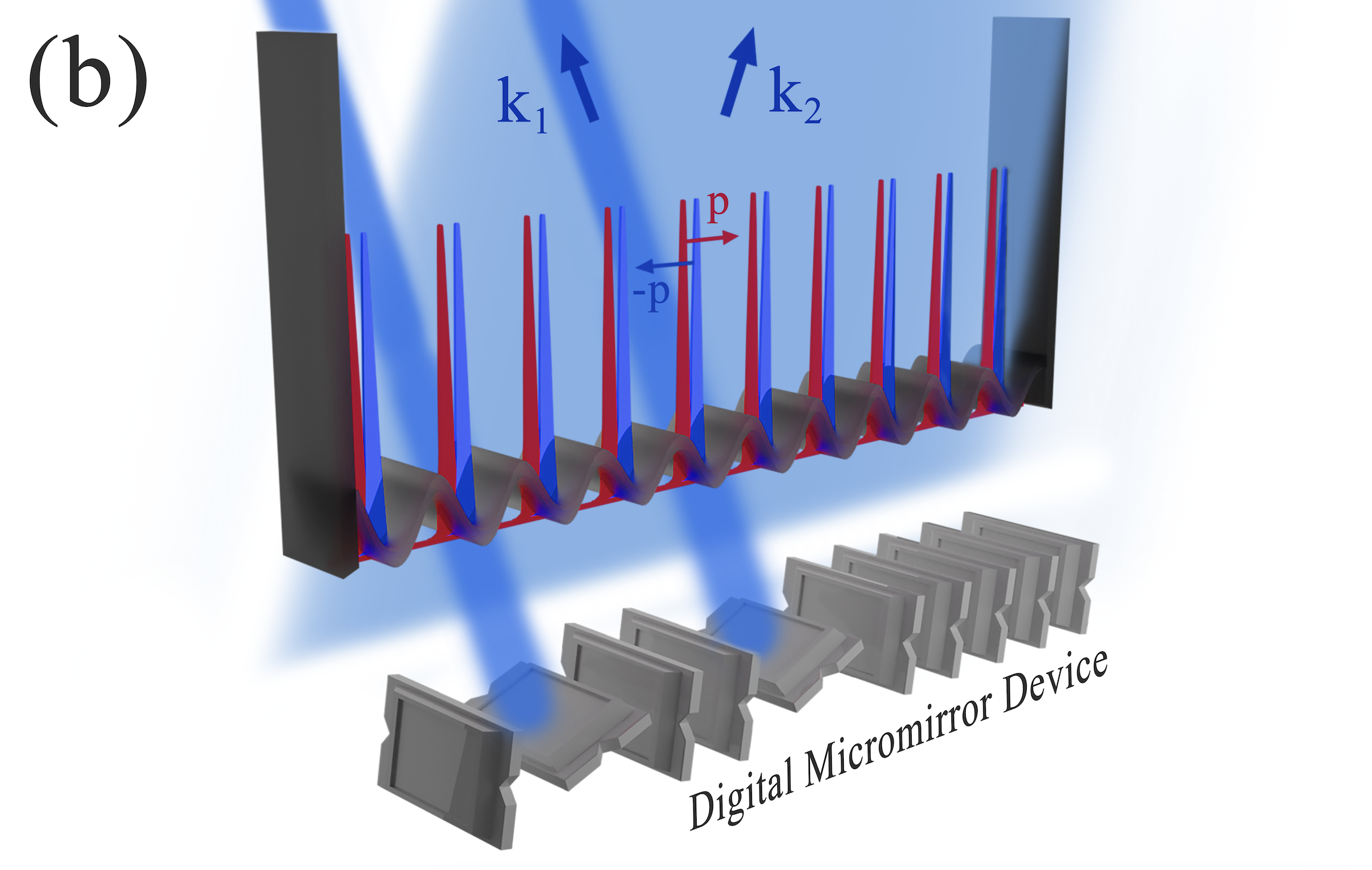}
\hfill
\includegraphics[width=0.329\textwidth]{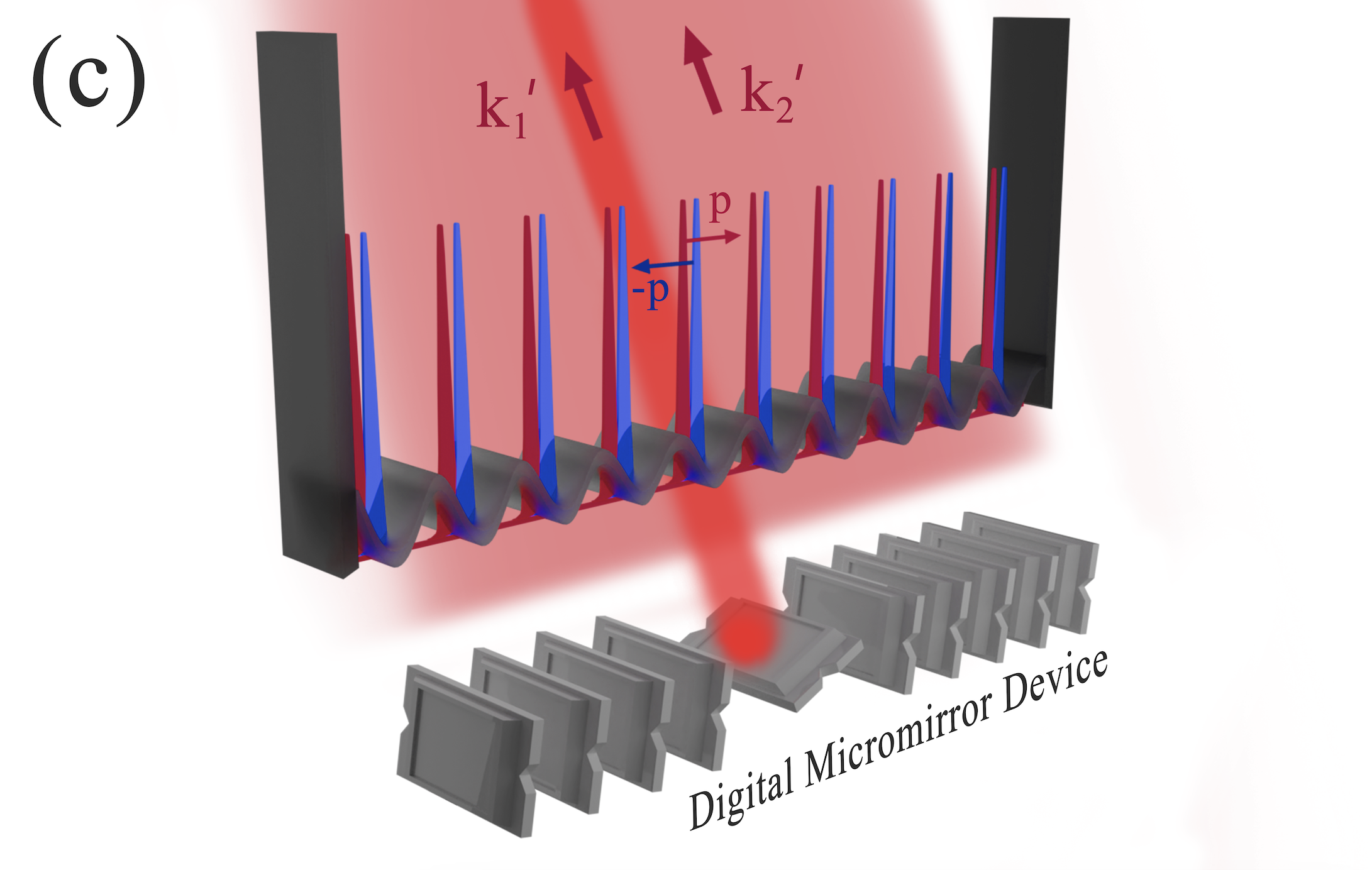}
\caption{
    {\bf Controlling tunneling and interactions between resonantly driven quantum wave-packets.}
    {\bf (a)}:~Red ball illustrates a classical particle moving in a 1D box potential in the presence of an oscillating spatially periodic potential. If the period of particle motion is $s$ times longer than the period of the oscillating potential, $T=2\pi/\omega$, we have an $s:1$ resonance. In such a case, $s$ classical non-interacting particles can be positioned so that they follow one another along the resonant trajectory. In the quantum description, the $s:1$ resonance manifests itself as $s$ localized wave-packets, which evolve one after another along the resonant trajectory. Here, we assume that the wave-packets are sufficiently localized so that {\it natural} tunneling between neighboring wave-packets is negligible. {\bf (b)}:~If during the encounter of the $i$-th and $j$-th wave-packets moving in opposite directions, one broad laser beam (labelled $\vect k_2$) and one narrow laser beam (labelled $\vect k_1$) focused on the meeting point of the wave-packets are briefly turned on and the Bragg scattering \cite{Muller2008} condition is satisfied, we can realize atom tunneling between the wave-packets with an amplitude $J_{ij}$, the modulus of which depends on the beam parameters and the interaction time with the atom, and a phase depending on the relative phase between the beams. With a controllable array of focused laser beams available (e.g., using a digital micromirror device \cite{Wang2020DMD} as in the figure), which can be independently activated at different time points, one can control the tunneling amplitudes $J_{ij}$ between any pair of wave-packets. {\bf (c)}:~Ultra-cold atoms can be prepared via a Feshbach resonance in states where they do not interact. In such a situation, we have the possibility of selectively controlling interactions between atoms occupying any pair of wave-packets. If just before the encounter of atoms occupying two wave-packets, we perform a Raman transfer \cite{Levine2022} (using broad and focused laser beams) from the initial internal states of the atoms to states where the atoms interact, then during the passage of the wave-packets, interaction between atoms will occur. After the wave-packets pass each other, we perform a Raman transfer again, but back to non-interacting internal states of the atoms. With a digital micromirror device available, interaction between atoms occupying any pair of wave-packets can be realized.}
\label{fig1}
\end{figure*}

Consider a single atom confined within a one-dimensional (1D) box potential periodically perturbed by an oscillating optical lattice potential [Fig.~\ref{fig1}(a)]. Using units of length, energy, and time, denoted by $1/k$, $\hbar^2 k^2/m$, and $m/\hbar k^2$, respectively, where $k$ is the wave number of the laser beam creating the oscillating optical lattice potential, and $m$ is the mass of the atom, the system Hamiltonian is given by
\be
H_0=\frac{p^2}{2}+\frac{\lambda}{2} \cos^2(z)+\frac{\lambda}{2} \cos^2(z)\cos(\omega t),
\label{H0}
\ee
where $\lambda$ and $\omega$ represent the amplitude and frequency of the oscillating optical lattice potential, and $z$ ranges from 0 to the size of the box, $L = s\pi/2$, where $s$ is an integer. In the classical description, if the driving frequency is $s$ times larger than the frequency of the unperturbed motion of the atom, i.e., $\omega = s|p|\pi/L=2|p|$, where the resonant momentum $|p|\gg\sqrt{\lambda}$, the condition for the $s:1$ resonance is met \cite{Buchleitner2002}. For a given optical lattice potential (i.e., for a selected wavelength of the laser beam, $2\pi/k$), the resonant number $s$ depends on the size of the 1D box potential. By changing the size of the 1D box, we can achieve any value of $s$.

Before considering the quantum resonant behavior of the atom, let us commence with a classical description. We introduce a new pair of momentum and position variables, known as action-angle variables \cite{Buchleitner2002}, $I = s|p|/2$ and $|\theta| = 2z/s$ where $-\pi \leq \theta < \pi$. Switching to the moving frame, $\Theta = \theta - \omega t/s$, and neglecting rapidly oscillating terms, the effective Hamiltonian becomes $H_{\rm eff} = P^2/2 + (\lambda s^2/32)\cos(s\Theta)$, where $P = I - \omega/s$ \cite{Buchleitner2002,SachaTC2020}. For $s \gg 1$, $H_{\rm eff}$ describes an atom moving in a spatially periodic potential. In the quantum description, this results in solid-state-like behavior, which is manifested as the formation of energy bands and eigenstates in the form of Bloch waves \cite{SachaTC2020,GuoBook2021}.

Upon returning from the moving frame to the laboratory frame, this solid-state-like behavior is observed in the time domain \cite{SachaTC2020}. Indeed, when fixing the position in space ($\theta = \rm const$) in the laboratory frame and examining how the probability of observing the atom at a chosen point changes in time, this replicates the crystalline structure observed versus $\Theta$ in the moving frame. This is due to the linear time transformation between the laboratory and moving frames, i.e., $\Theta = \theta - \omega t/s$. This concept has facilitated the exploration of various condensed matter phases in the time domain across different resonantly driven single-particle and many-body systems \cite{SachaTC2020,Hannaford2022}.

In the following, we confine our focus to the lowest energy band of the quantum version of the Hamiltonian $H_{\rm eff}$. The corresponding Hilbert subspace can be expanded in the Wannier state basis, $w_i(\Theta)$, where $w_i(\Theta)$ is localized at the $i$-th site of the potential in $H_{\rm eff}$ \cite{SachaTC2020,Hannaford2022}. In the laboratory frame, these are localized wave-packets, $w_i(z,t)$, evolving along the periodic resonant trajectory with the resonant momentum $p$, i.e., with the period $sT$, where $T=2\pi/\omega$ [Fig.~\ref{fig1}(a)]. In this basis, the Hamiltonian of many non-interacting bosonic atoms takes the form
\be
\hat H=\frac{J}{2}\sum_{i=1}^s\hat a_{i+1}^\dagger \hat a_i+\rm{h.c.},
\ee
where $\hat a_i$ ($\hat a_i^\dagger$) represents the bosonic annihilation (creation) operator, responsible for annihilating (creating) a particle in the $i$-th wave-packet and
$J=2\int_0^Ldz,w_{i+1}^*(H_0-i\partial_t)w_i$ represents the tunneling amplitude of atoms between the neighboring wave-packets on the resonant trajectory \cite{SachaTC2020,Wang2021}. We set the oscillation amplitude $\lambda$ such that these {\it natural} tunnelings are negligible ($J\approx 0$), as our intention is to selectively introduce tunneling 
in a controlled manner.

\subsection{Control of the tunneling amplitudes}

To selectively control atom transfer between different wave-packets, we can utilize Bragg scattering \cite{Muller2008}. Let us consider two laser beams characterized by wave vectors $\vect k_1$ and $\vect k_2$ that we switch on for a short period of time $\tau$. The first beam is focused at the location where two wave-packets, $w_i(z,t)$ and $w_j(z,t)$, meet during evolution [Fig.~\ref{fig1}(b)]. The waist of the beam is larger than the width of the wave-packets but smaller than the distance between potential minima in (\ref{H0}). The second beam is broad and covers the entire 1D box. If the Bragg condition is satisfied, $\vect k_1-\vect k_2=2p\vect e_z$, an atom occupying the wave-packet $w_j$, moving with an average momentum $p\vect e_z$, is transferred to the other wave-packet $w_i$, which moves with an average momentum $-p\vect e_z$, and the reverse transfer is equally probable. These processes are described by an additional term in the Hamiltonian \cite{Muller2008} 
\be 
H_B=\lambda_{\rm Bragg}(t)\cos^2\left[\frac{(\vect k_1-\vect k_2)\cdot\vect e_z z+\phi}{2}\right] \frac{e^{-\frac{(z-z_0)^2}{W^2}}}{\sqrt{4\pi W^2}}, 
\label{hbragg}
\ee
where $z_0$ and $W$ are the location and waist of the narrow laser beam, respectively. We choose $W=\pi/(2\sqrt{2})$. $\lambda_{\rm Bragg}$ is determined by the intensities of the beams and $\phi$ is the relative phase of the beams. The beams are activated only for a short period of time $\tau$ around the moment when two relevant wave-packets pass each other. The resulting tunneling amplitude of the atom between the wave-packets is described by 
\begin{equation}
J_{ij}(t)=2\int_0^L dz \;w_i^*(z,t)\;H_B(z,t)\;w_j(z,t). 
\end{equation}
The magnitude of $J_{ij}$ is controlled by the parameters of the beams and the duration of their interaction with the atom (in the following we choose $\tau=0.24T$, where $T=2\pi/\omega$, around central moments of wave-packet encounters), while the phase is controlled by the relative phase $\phi$. By switching on the beams during meetings of different wave-packets and having multiple focused beams available [Fig.~\ref{fig1}(b)], we have the ability to control the transfer of atoms between all wave-packets, i.e., we can realize the Hamiltonian
\be
\hat H=\frac12\sum_{i,j=1}^sJ_{ij}(t)\;\hat a_i^\dagger \hat a_j.
\label{t-b}
\ee

\subsection{Control of the interaction strengths}

Interactions between ultra-cold atoms are effectively short-range contact interactions described by a single parameter, the s-wave scattering length \cite{Pethick2002}. In the Hilbert subspace spanned by the wave-packets, $w_i(z,t)$, evolving along the periodic resonant trajectory, the Hamiltonian of interacting atoms reads  \cite{SachaTC2020,Wang2021}
\be
\hat H(t)=\frac12\sum_{i,j=1}^s\left(J_{ij}(t)\;\hat a_i^\dagger \hat a_j+U_{ij}(t)\;\hat a_i^\dagger \hat a_j^\dagger \hat a_j \hat a_i\right),
\label{Huniversal}
\ee
with the interaction coefficients,
\be
U_{ij}(t)=g_0(t)\int_0^Ldz\;|w_i(z,t)|^2\;|w_j(z,t)|^2, 
\label{uij}
\ee
where $g_0(t)=2m\omega_\perp a(t)/\hbar k$ and $a(t)$ is the scattering length which can be changed in the experiment. Choosing properly an external magnetic field in the vicinity of a Feshbach resonance \cite{Pethick2002}, we can set the scattering length $a=0$. If we change the internal state of the atoms that occupy two wave-packets which are about to meet during the evolution along the periodic trajectory, they will interact because the scattering length $a$ is no longer zero. The change of internal state can be performed by a focused laser beam which together with another broad beam realizes a Raman transition between the internal states \cite{Levine2022}, see Fig.~\ref{fig1}(c). Immediately after the wave-packets pass each other, the laser beams can be applied once again to restore the initial internal states of the atoms. The entire procedure leads to a time-dependent $g_0(t)$ which is non-zero only during the passing of the wave-packets. With an array of focused laser beams one can control and engineer all interaction coefficients $U_{ij}$ in the Hamiltonian (\ref{Huniversal}).

\subsection{The idea of a temporal printed circuit board}

We have demonstrated that ultra-cold atoms confined in a box potential and driven by a periodically oscillating optical lattice can be described by the Hamiltonian (\ref{Huniversal}). In this framework, it is possible to engineer and precisely control tunneling amplitudes, $J_{ij}$, between arbitrary sites via Bragg scattering, as well as tune the interaction strengths, $U_{ij}$, between atoms occupying any selected pair of sites and everything can be modified at any moment during the experiment. The Hamiltonian (\ref{Huniversal}) is derived for bosons, but a similar universal Hamiltonian can be obtained for fermions. 

\begin{figure*}[t]
\begin{minipage}{0.45\textwidth}
\includegraphics[width=1.0\textwidth]{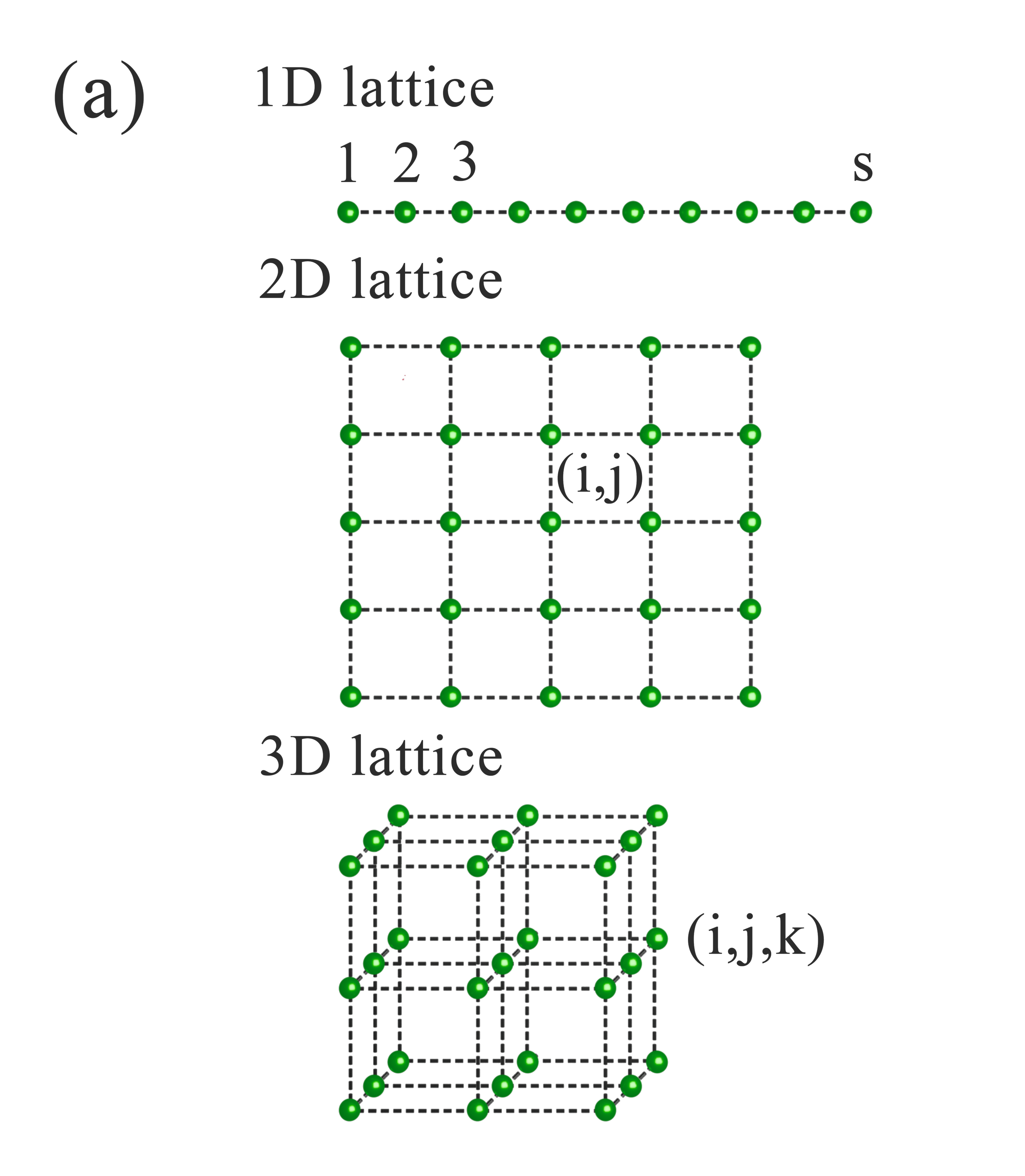}\\
\includegraphics[width=1.0\textwidth]{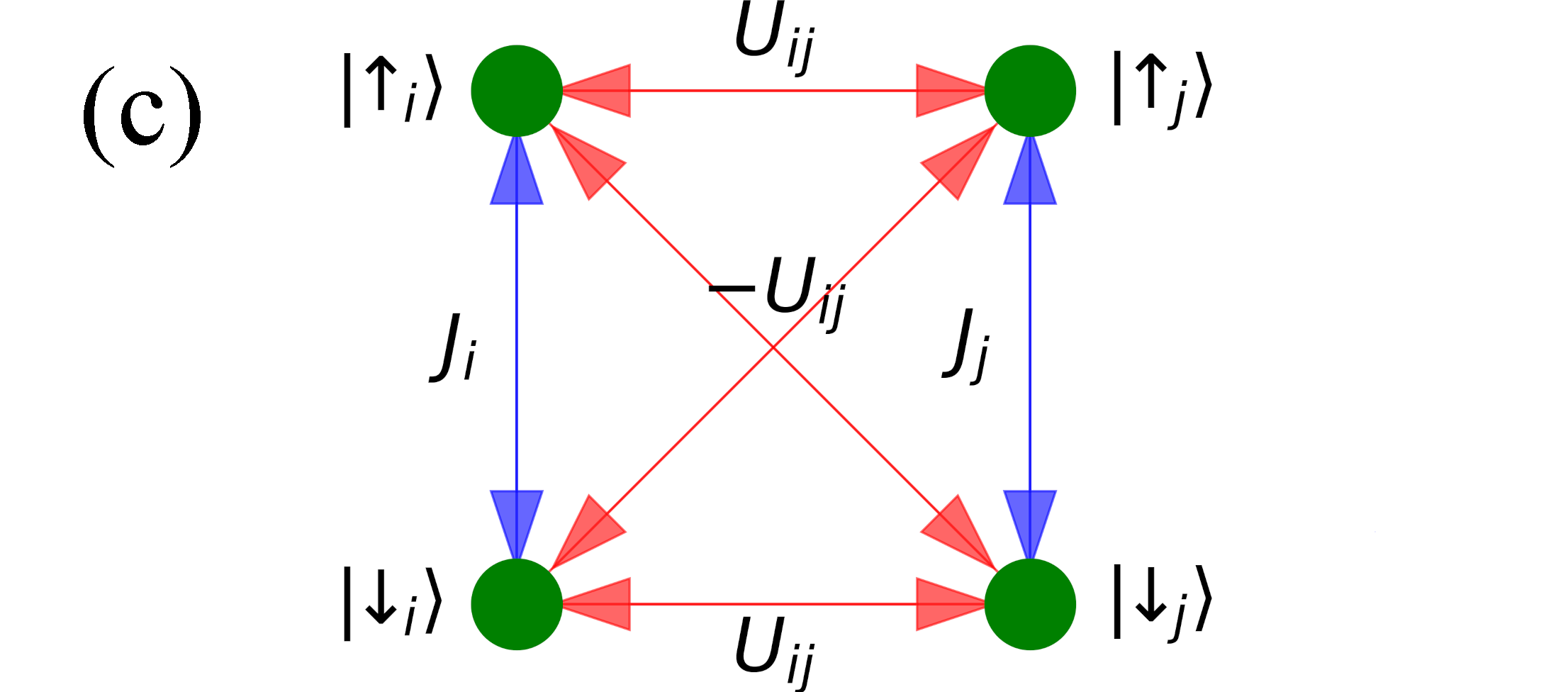}
\end{minipage}
\begin{minipage}{0.54\textwidth}
\includegraphics[width=0.65\textwidth]{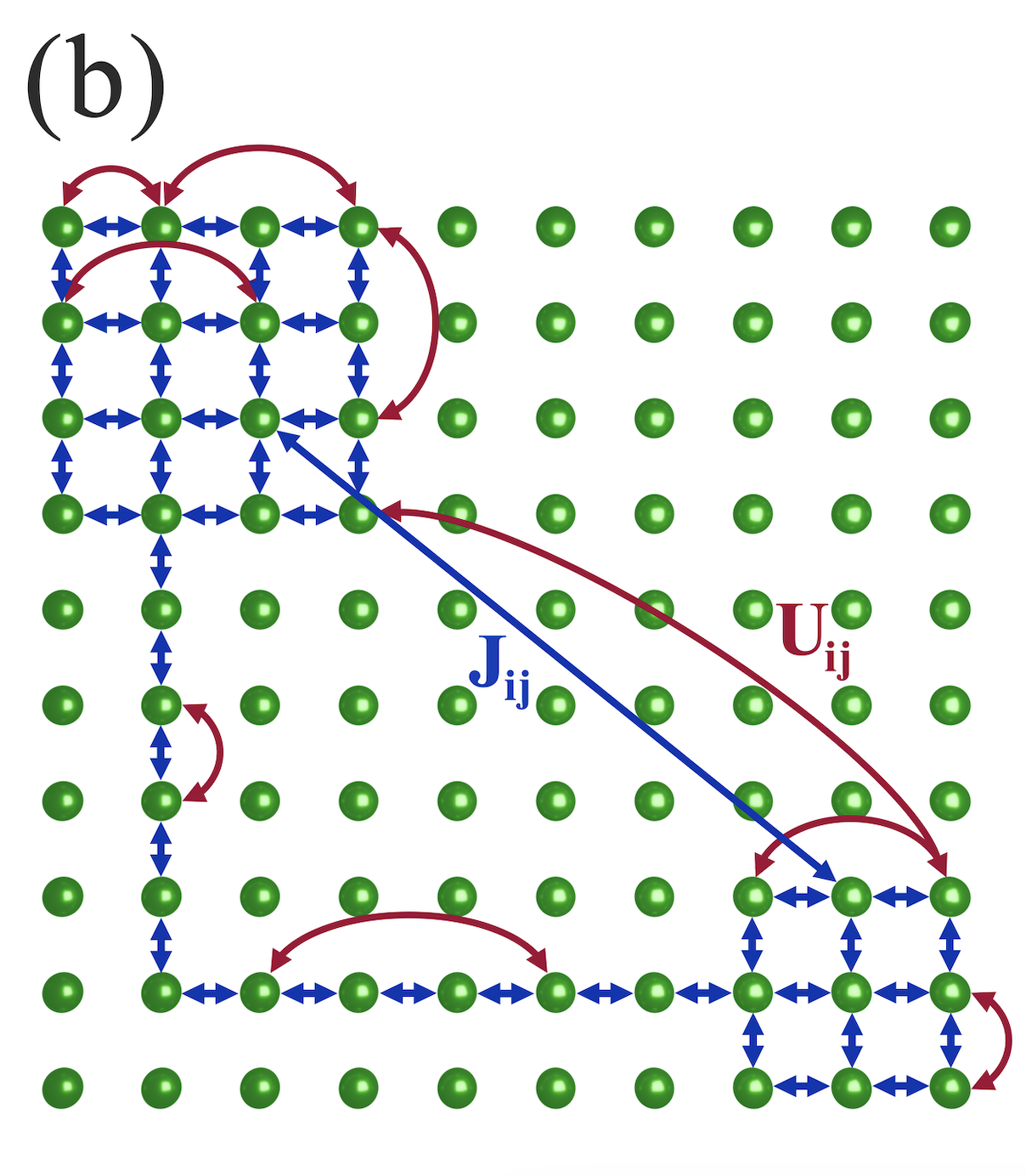}
\includegraphics[width=0.65\textwidth]{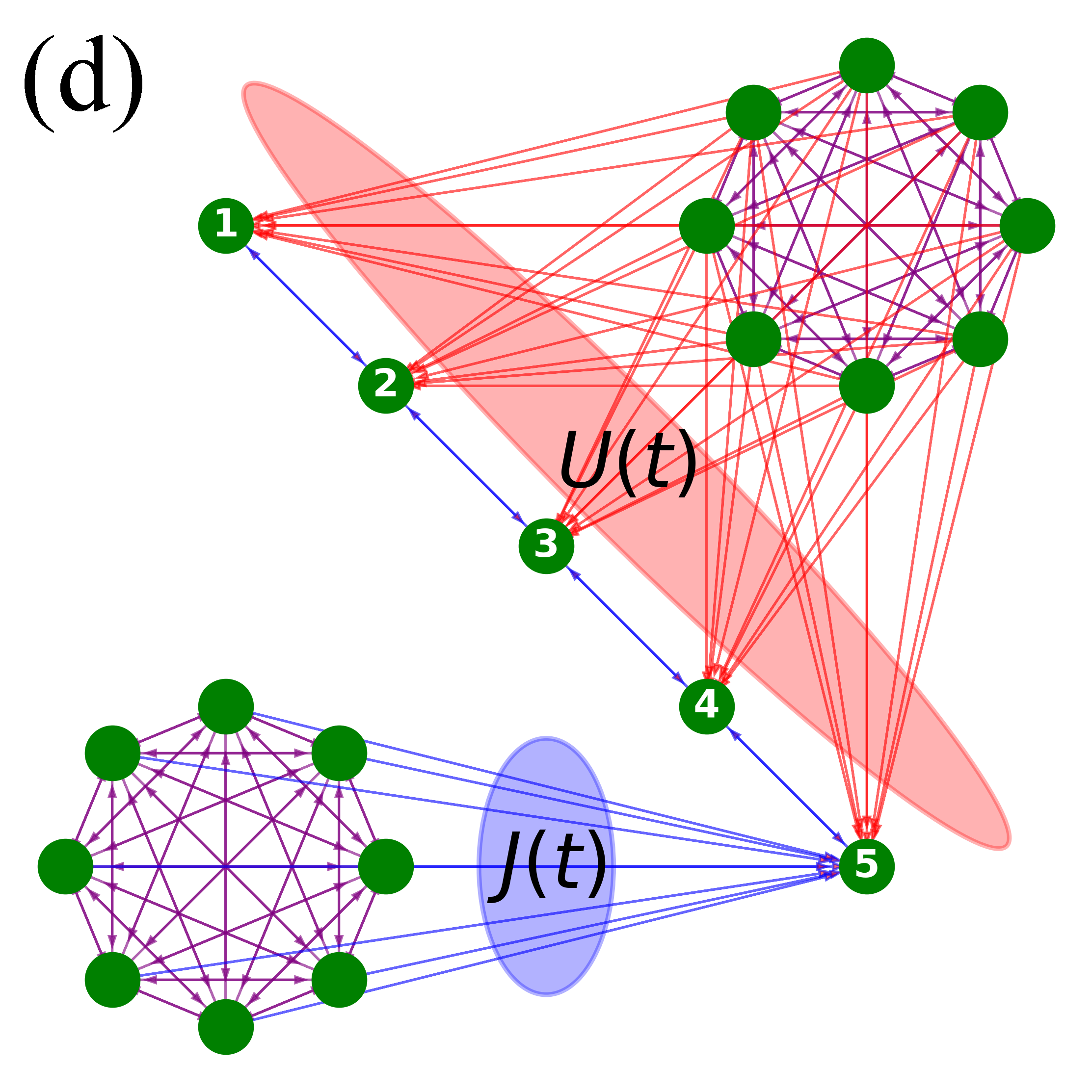}
\end{minipage}
\caption{
{\bf Temporal printed circuit board. (a)}: The $s$ wave-packets evolving along the resonant trajectory [Fig.~\ref{fig1}(a)] can be treated as $s$ states of a lattice with $s$ sites. If we arrange the sites of the lattice in a 1D chain and using Bragg scattering [Fig.~\ref{fig1}(b)] we realize tunneling between nearest neighbors, we will have a 1D crystalline structure with nearest neighbor hoppings. If we arrange the sites in a 2D lattice and induce tunneling between nearest neighbors, we will have a 2D crystalline structure. Similarly, we can realize 3D and higher-dimensional crystalline structures. {\bf (b)}: Let us consider $s$ periodically evolving wave-packets as states of a 2D lattice. Using a broad laser beam and having only two focused laser beams at our disposal [Fig.~\ref{fig1}(b)], all nearest neighbor tunnelings of the 2D lattice can be realized. For example, a 2D sublattice can be coupled to another 2D sublattice via a 1D chain as shown in the figure (realized tunnelings are marked with short blue arrows). Control over the phases of tunneling amplitudes $J_{ij}$ allows for the generation of an artificial magnetic field in the system. With an array of focused laser beams available, any sites in the lattice can be connected via tunneling, as illustrated by the long blue arrow. More exotic geometries can also be realized, such as the Klein bottle, where the left and right edges of the 2D lattice are connected normally, but the top edge is twisted before connection to the bottom edge. Using selective Raman transfer [Fig.~\ref{fig1}(c)], interactions between atoms occupying any pair of sites can be realized and controlled (as indicated by red arched arrows in the plot). 
{\bf (c)}: Example building block of two connected $\operatorname{SU}(2)$ systems. The number of bosonic atoms decides the representation in each system $i$ --- if there is one atom, we deal with the spin-1/2 system. Blue vertical arrows correspond to tunneling (ladder operators), the (anti-) ferromagnetic interactions can be supplied by programmed interactions depicted by horizontal or diagonal lines.
{\bf (d)}: Example of 1D Bose-Hubbard system of 5 sites (labelled 1 to 5 in the plot) connected to two engineered reservoir systems. The reservoirs represented by all-to-all connected systems allow for extremely fast mixing and their spectral density can be designed. During an experiment, the parameters can undergo arbitrary reconfigurations. Here, the purely dephasing interactions $U(t)$ with one reservoir and particle exchange with the other reservoir $J(t)$ can be, e.g., periodically switched during the experiment allowing for study of quantum thermodynamics in a fully closed system.
}
\label{fig2}
\end{figure*}

With the universal time crystalline structure described by the Hamiltonian (\ref{Huniversal}), it is possible to realize a broad range of quantum devices. Figure~\ref{fig2} presents a few examples. Physically, the states associated with the sites of the time crystalline lattice are localized wave-packets moving periodically in a 1D box [Fig.~\ref{fig1}(a)]. One can represent the crystalline structure as a row of sites but if the sites are connected via selective Bragg tunneling as depicted in the middle plot in Fig.~\ref{fig2}(a), then the structure can be represented as a 2D lattice with nearest neighbor tunnelings, and 2D condensed matter problems can be realized and investigated. For example, bosonic or fermionic atoms  can experience a magnetic-like field if proper complex phases of the tunneling amplitudes $J_{ij}$ are realized by a proper choice of the relative phase, $\phi$, of the laser beams in the Bragg scattering, see (\ref{hbragg}). A few such 2D lattices can be created and connected via selective tunneling to form a 3D lattice [bottom plot in Fig.~\ref{fig2}(a)], and this procedure can be continued to form a 4D lattice, and so on. It is also possible to realize exotic objects like a Klein bottle, which is hard to imagine in 3D space \cite{Boada2015,Grass2025}. That is, pairs of opposite edges of a 2D lattice can be connected via tunneling, but one of the edges is twisted before the connection.

An entire time crystalline structure can be represented as sites on a 2D board, as shown in Fig.~\ref{fig2}(b). Such a board can be considered analogous to a printed circuit board in electronics. Indeed, any site on the board can be connected to any other site through selective tunneling, and an atom occupying any site can interact with an atom occupying any other site. All tunnelings and interactions can be individually turned on or off or modified because all parameters of the Hamiltonian (\ref{Huniversal}) are fully controlled. Thus, in one part of the board, we can realize, for example, a 2D structure, while in other parts of the board, there are other 2D or higher-dimensional structures, or a structure with fully connected sites, or even a structure in the form of a Klein bottle. We can realize and control arbitrary connections between the structures and reconfigure the entire system at any moment during the experiment. Consequently, we can realize a broad range of quantum devices that can perform the operations we need --- examples are presented in Figs.~\ref{fig2}(c)-\ref{fig2}(d).

Since the coupling coefficients $J_{ij}(t)$ and $U_{ij}(t)$ are activated at certain moments of time, the entire Hamiltonian (\ref{Huniversal}) is time-dependent. However, if we apply a sequence of the Bragg and Raman pulses in the same way every period of the resonant trajectory and if 
\begin{equation}
    \frac{1}{|J_{ij}|}\gg sT,\quad \frac{1}{|U_{ij}|N/s}\gg sT,
\end{equation}
we may derive the effective time-independent Hamiltonian. That is, if the timescales associated with atomic tunneling and interactions between atoms are much longer than the period of the resonant trajectory, it is justified to perform a time average of the original Hamiltonian, yielding the effective Hamiltonian
\begin{equation}
    \hat H_{\rm eff}=\frac12\sum_{i,j=1}^s\left(\bar J_{ij}\;\hat a_i^\dagger \hat a_j+\bar U_{ij}\;\hat a_i^\dagger \hat a_j^\dagger \hat a_j \hat a_i\right),
\end{equation}
where 
\begin{eqnarray}
    \bar J_{ij}&=&\frac{1}{sT}\int_0^{sT}dt\; J_{ij}(t), \\
    \bar U_{ij}&=&\frac{1}{sT}\int_0^{sT}dt\; U_{ij}(t).
\end{eqnarray}
The circuit board is now described by a time-independent Hamiltonian $\hat H_{\rm eff}$, yet at any moment during the experiment we can fully reprogram the sequence of Bragg and Raman pulses, thereby driving the system to evolve within an entirely different effective structure. This enables the realization of a sequence of different quantum operations on a system of ultra-cold atoms. The initial operations may be used to prepare an initial state of the many-body system, while subsequent ones can perform the desired computational tasks.

We describe the realization of the temporal printed circuit board in ultra-cold atoms moving between the walls of a box potential. This potential offers significant advantages, as the laser beam parameters required for Bragg scattering can remain identical regardless of where the evolving wave-packets meet. Box potentials are implemented in ultra-cold atomic gas laboratories and are used in experimental studies of many-body physics \cite{Gaunt2013}. However, we emphasize that temporal printed circuit boards can also be realized in other trapping potentials and also for atoms bouncing on an atomic mirror \cite{Giergiel2020}. Any trapping potential that is not harmonic allows for stable resonant atomic motion. Similar to the case of the box potential, the same tools (Bragg scattering and Raman transitions) can be used to implement the temporal printed circuit board. The only difference is that atoms evolving along a resonant trajectory will have different momenta at different positions due to variations in potential energy. Consequently, the parameters of Bragg pulses must be adjusted at different points where evolving wave-packets meet. 

\begin{figure*}[t]
\includegraphics[width=0.49\textwidth]{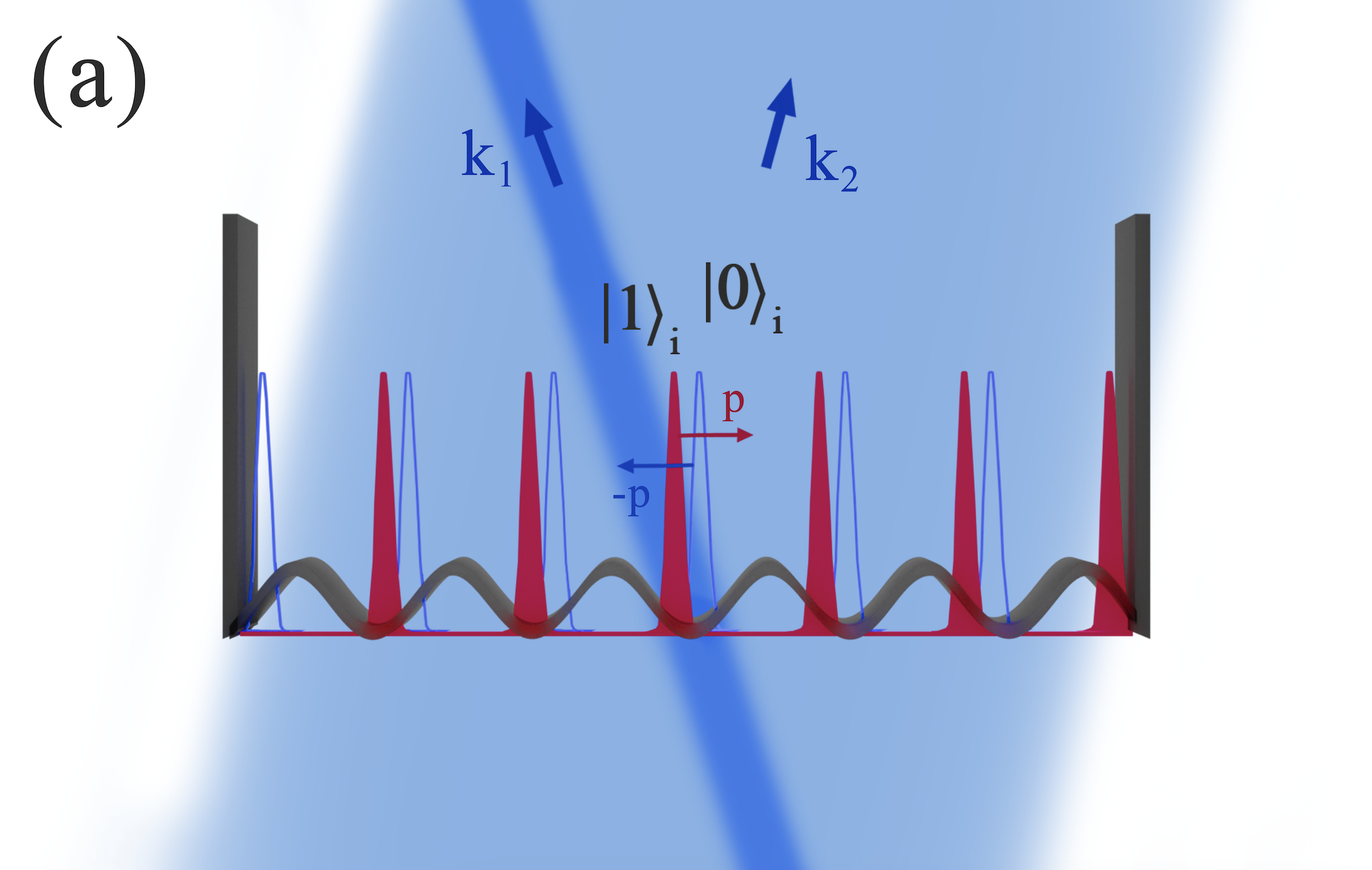}
\hfill
\includegraphics[width=0.49\textwidth]{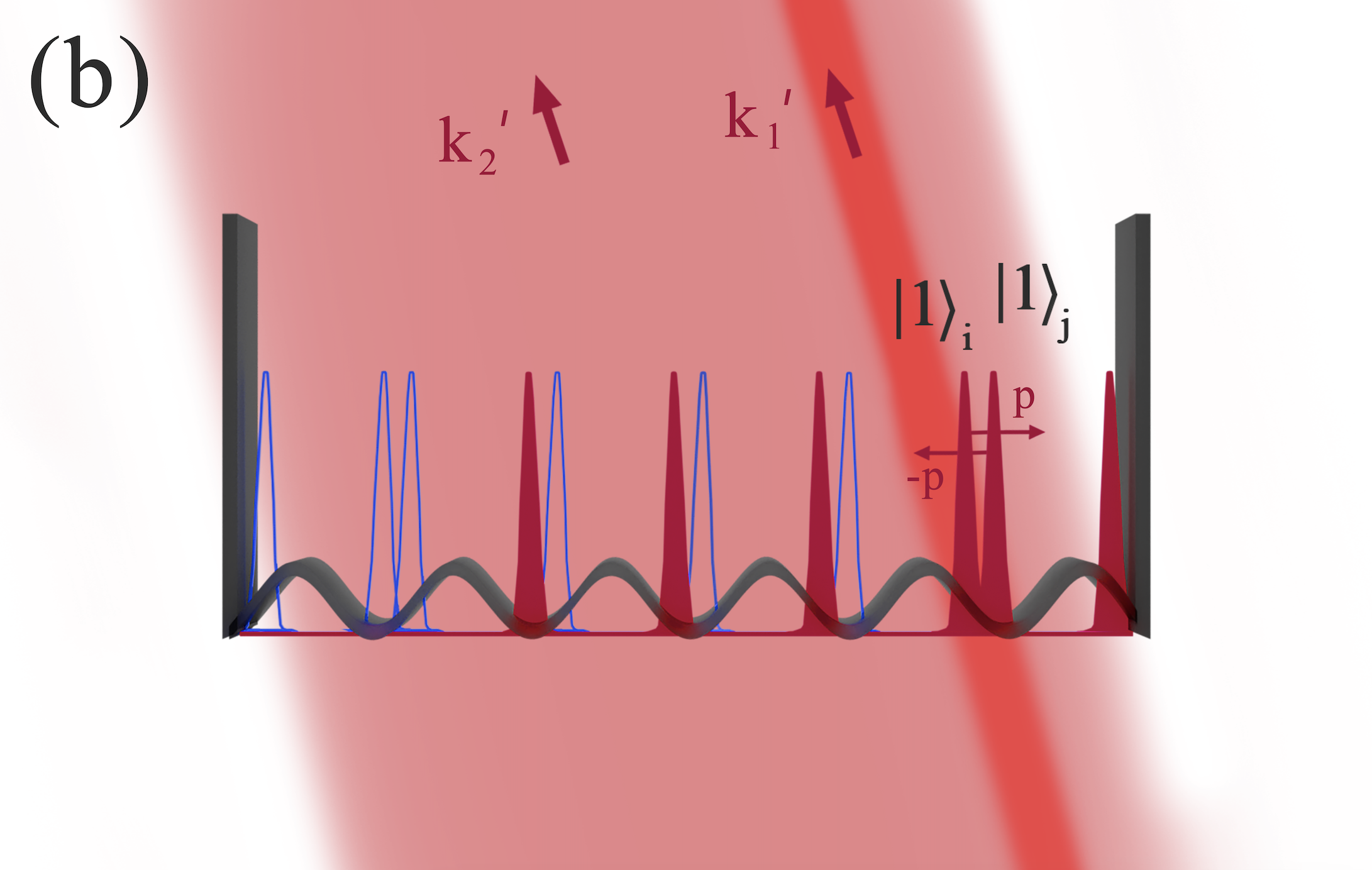}
\caption{
{\bf Quantum computer. (a)}: Initially, $s/2$ bosonic atoms are prepared in a Mott insulator phase in a 1D box potential in the presence of a static optical lattice. We assume there are $s/2$ potential wells in the box, with one atom in each well. After preparing the Mott insulator phase, the interactions between atoms are turned off by means of a Feshbach resonance \cite{Pethick2002}. The next step is to impart momentum to the atoms that satisfies the $s:1$ resonance condition with the frequency of the optical lattice oscillation, which we simultaneously switch on. Shortly after the previously described procedure, we observe $s/2$ wave-packets occupied by single atoms moving to the right in the figure and $s/2$ unoccupied wave-packets moving to the left. We assign one unoccupied wave-packet to each occupied wave-packet, forming the $|0\ra$ and $|1\ra$ states of qubits. In total, we have $s/2$ qubits. When two wave-packets corresponding to the $|0\ra$ and $|1\ra$ states of the same qubit meet during evolution along the resonant trajectory, a single-qubit gate can be performed using Bragg scattering. Control over the relative phase of the laser beams used in the Bragg scattering allows us to control whether the $\sigma_x$ or $\sigma_y$ operation is performed. High fidelity single-qubit operations are achieved by dividing them into several stages, meaning the entire single operation requires multiple encounters of wave-packets, and hence spanning several periods of the resonant trajectory, see Fig.~\ref{fig4}(a). {\bf (b)}: When wave-packets corresponding to the states $|1\ra_i$ and $|1\ra_j$ of the $i$-th and $j$-th qubits meet during evolution, a controlled-Z gate can be realized. Atoms initially do not interact, but just before the wave-packets pass each other, we change the internal states of the atoms to states where they interact. We do this using Raman transfer with two beams. After the wave-packets pass each other, another Raman transfer restores the initial internal states of the atoms. If the interaction strength between atoms and the duration of interaction are appropriately chosen, after the wave-packets pass, the state $|1\ra_i|1\ra_j$ acquires a phase $e^{i\pi}$, and the controlled-Z gate is completed. Similarly to single-qubit operations, a high fidelity controlled-Z gate can be realized by dividing it into several encounters of the appropriate wave-packets, see Figs.~\ref{fig4}(b)-\ref{fig4}(c). 
}
\label{fig3}
\end{figure*}

\section{Quantum computer}
\label{qcsec}

As a compelling demonstration of the versatility of our temporal printed circuit board, we explore its application to the implementation of a quantum computer. Although not the central focus of this work, this represents one of the most experimentally challenging architectures, making it an ideal and stringent benchmark for showcasing both the power and the fundamental limits of our approach.

\subsection{The idea of realizing a quantum computer}
\label{ideaqc}

The construction of a universal quantum computer requires the preparation of numerous qubits (e.g., two-level atoms), the ability to perform all single-qubit operations, and the capability to execute two-qubit operations (e.g., controlled-Z gate operation) between any pairs of qubits \cite{Preskill2018}. The latter necessitates selective interactions between atoms, which, in turn, demand precise coherent transport of atoms and activation of interactions when they are in close proximity \cite{Cirac2000,Bluvstein2022,Gonzalez-Cuadra2023}.

Crystalline structures in time automatically address the atom transport problem, as atoms are prepared in wave-packets evolving along a periodic trajectory, and each atom individually encounters every other atom at some point in time. Additionally, every atom moving on the same trajectory means they experience the same spatially inhomogenous fields, thus likely reducing sources of phase noise. To realize a controlled-Z gate for any pair of atoms, it is sufficient to activate interactions between them at the moment of their encounter. The temporal printed circuit board described in the previous section is well-suited for this purpose. The board also enables the implementation of all necessary single-qubit operations required for building a quantum computer. For the successful construction of a quantum computer, all gate operations need to be as straightforward as possible. In the proposed approach, which we describe in the following, we only need a static arrangement of focused laser beams that will be activated at the appropriate moments, allowing the implementation of all necessary gate operations.

The crystalline structure in time, as described in Fig.~\ref{fig1}, consists of $s$ states ($s$ localized wave-packets). In such a structure, if we prepare $s/2$ atoms, we can define $s/2$ qubits, assigning two different wave-packets to each atom (see Fig.~\ref{fig3}). If the atom occupies the first chosen wave-packet, the qubit will be in the state $|1\rangle$ and if it occupies the second wave-packet, it will be in the state $|0\rangle$. The initialization process of all qubits in, for example, state $|1\rangle$ is relatively straightforward. In the 1D box potential and in the presence of a static deep optical lattice, we prepare bosons in a Mott insulator state with unit filling of lattice sites \cite{Pethick2002}, assuming the number of lattice sites between the walls of the box potential is $s/2$. Next, we turn off the interactions between atoms setting the s-wave scattering length to zero by means of a Feshbach resonance \cite{Pethick2002}. We then set the atoms moving in one direction with momentum $p$ that satisfies the $s:1$ resonance condition with the frequency of the optical lattice oscillation. To do this, we set the initially static optical lattice in motion so that in the moving frame the atoms follow the ground state of the moving potential well. This is achieved using a counter-diabatic driving process which involves a simple unitary transformation to the moving frame by applying a linear potential proportional to the acceleration with which the lattice is set in motion (see Appendix~\ref{appa} for details). We then activate the optical lattice oscillation. This leads to a situation where $s/2$ wave-packets of the time crystalline structure that are initially moving in the same direction are occupied by individual atoms, while the other $s/2$ wave-packets initially moving in the opposite direction are unoccupied, see Fig.~\ref{fig3}(a). For each occupied wave-packet, we assign any unoccupied wave-packet to form a qubit together. The choice of which unoccupied wave-packets are assigned to which occupied wave-packets is arbitrary.

\begin{figure*}[t]
\includegraphics[width=0.49\textwidth]{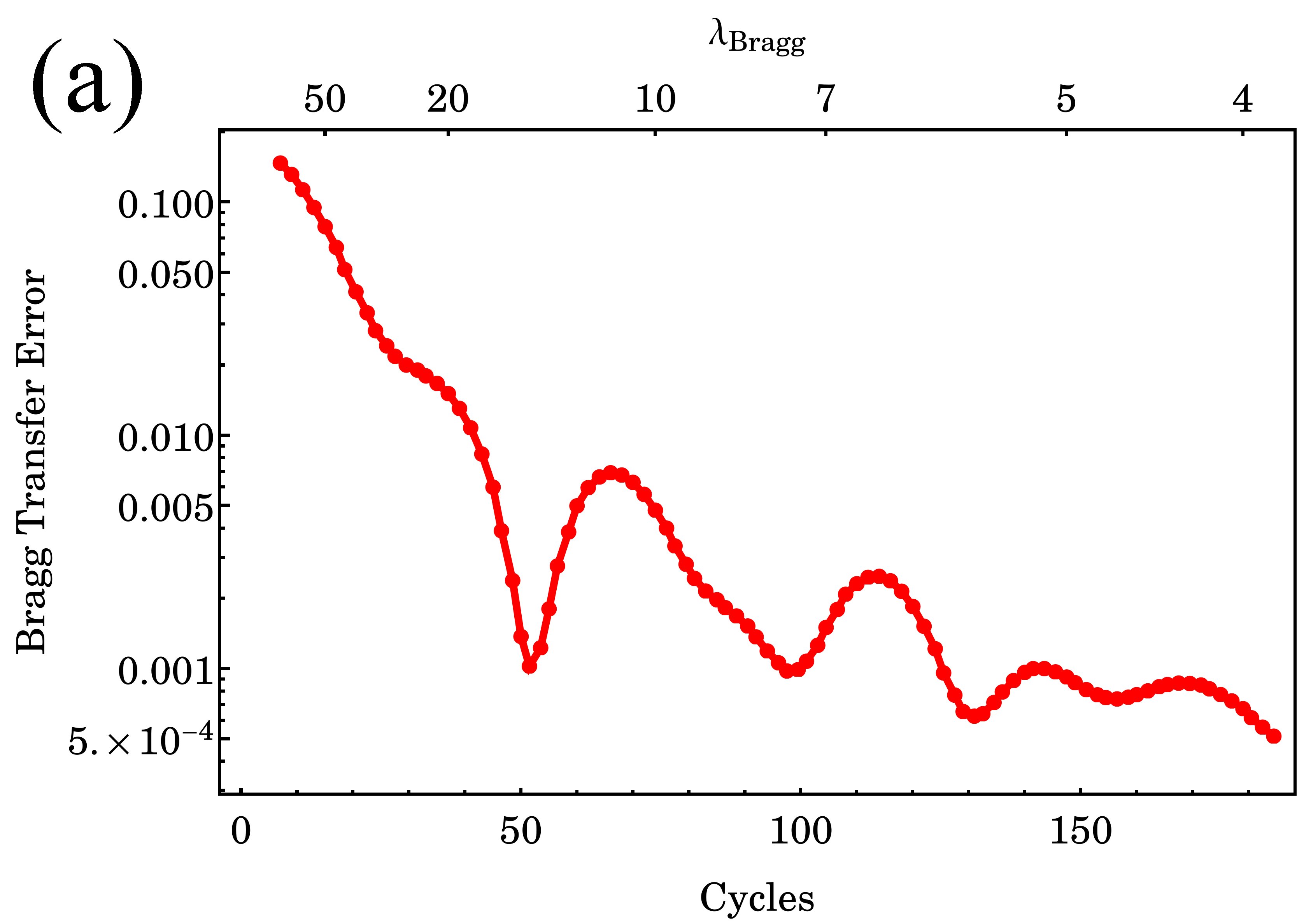}
\hfill
\includegraphics[width=0.49\textwidth]{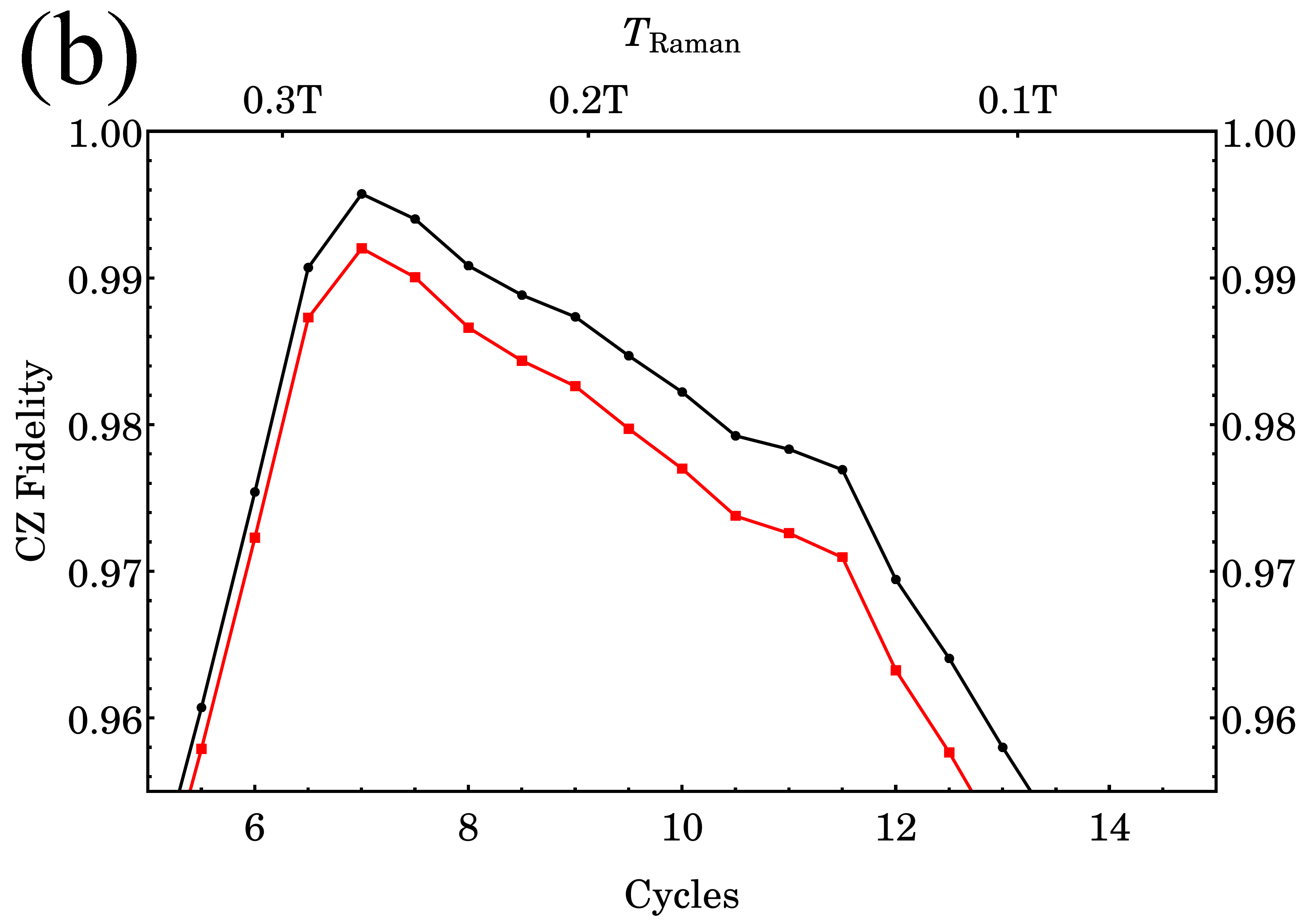}
\includegraphics[width=0.49\textwidth]{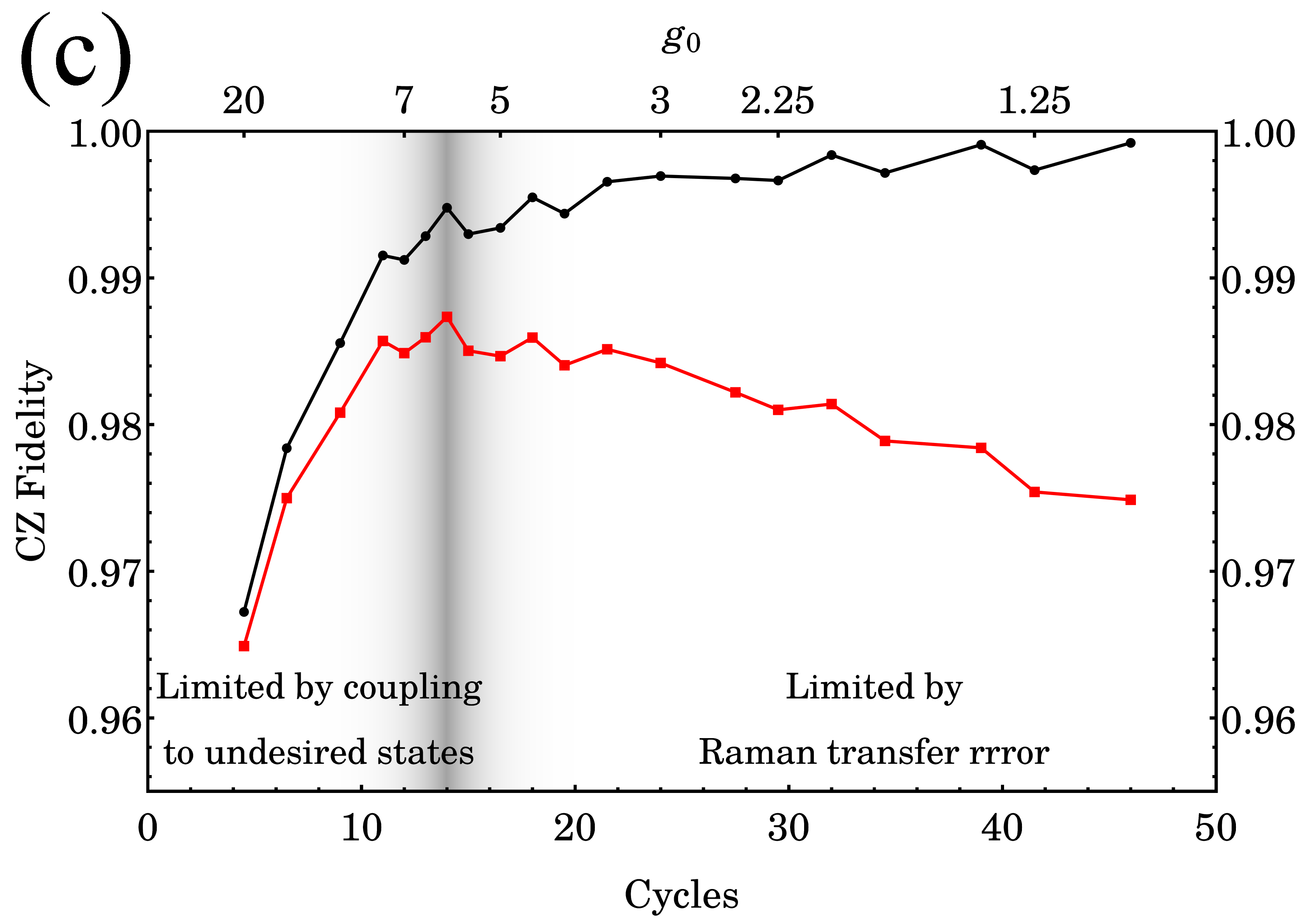}
\hfill
\includegraphics[width=0.49\textwidth]{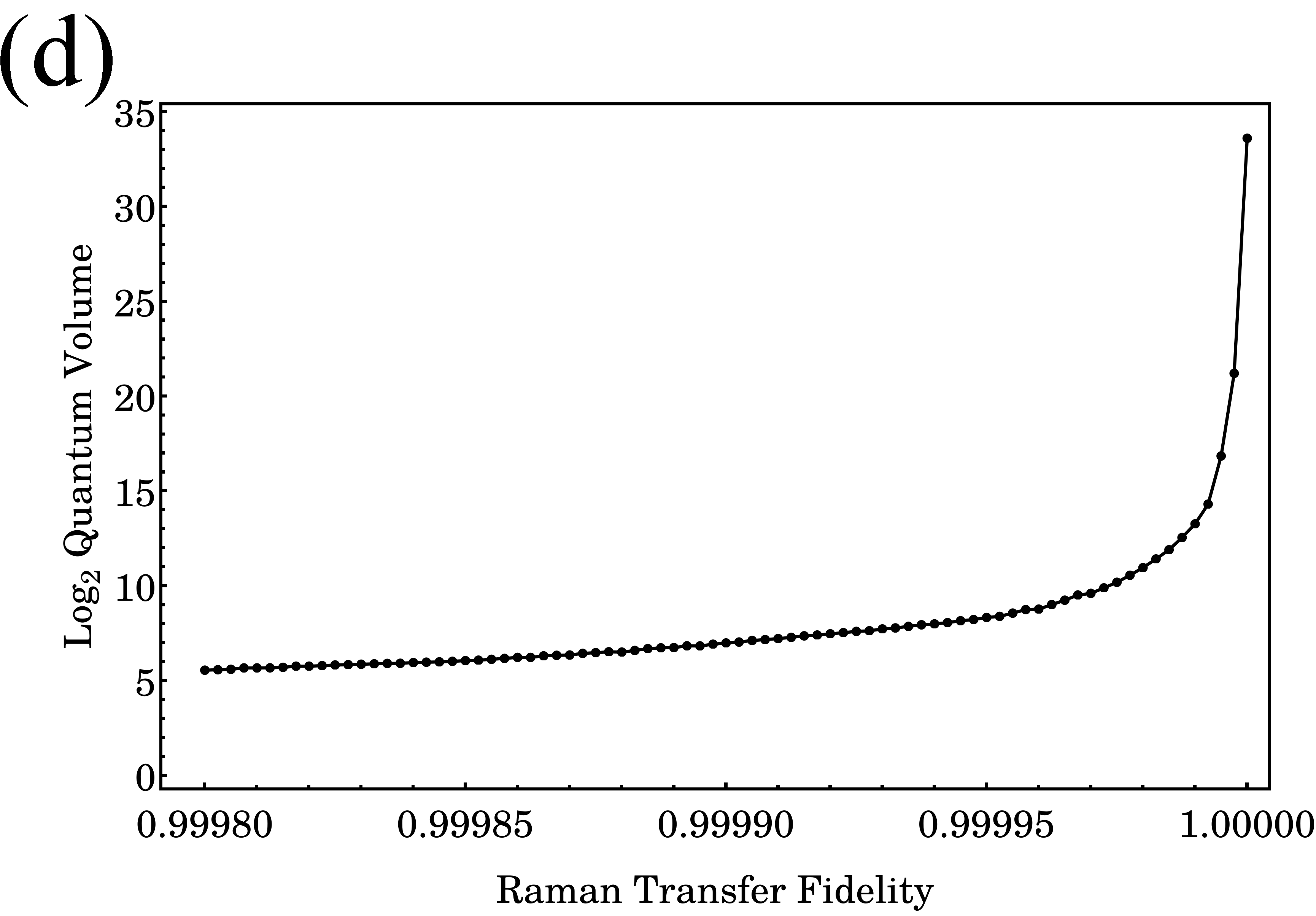}
\caption{
    {\bf (a)}:~Error of atom transfer between two wave-packets using Bragg pulses versus the assumed number of cycles of the resonant trajectory needed to achieve full transfer (two Bragg pulses per each cycle). Faster transfer requires stronger Bragg pulses, $\lambda_{\rm Bragg}$, increasing the coupling of the atom to other undesired states. Generally, apart from additional coherent oscillations, the longer the realization of the transfer, the smaller the error. {\bf (b)} and {\bf (c)}:~Fidelity of the controlled-Z gate (CZ), in which the interaction imparts a phase of $\pi$ to the state where one atom occupies one wave-packet and the second atom occupies another wave-packet, versus the number of required cycles (two interaction meetings of the wave-packets per cycle). The two curves represent the cases without (black) and with (red) the inclusion of Raman transfer errors \cite{Bluvstein2022}. For a small number of cycles, stronger interactions are needed, and the coupling to other states limits the CZ fidelity. For a larger number of cycles, more Raman transfers are needed, and their imperfections limit the fidelity. (b) is for a variable interaction $g_0$ and fixed Raman transfer time ($T_{\rm Raman} \approx 0.12T$, before and after the center-point meeting time) and (c) is for a fixed interaction ($g_0=10$) and variable Raman transfer time. 
   {\bf (d)}: Quantum volume \cite{Cross2018ValidatingQC} calculated based on the B-gate  decomposition of a generic SU(4) two-qubit operation \cite{Zhang2003MinimumCO}. In this plot we assume no error in the single-qubit operations. We plot a square of fidelilty of decomposition of a single B-gate using optimal points selected from the CZ-fidelity plot (b). For higher Raman transfer fidelity one can use a longer multi-cycle realization of the CZ-gates leading to quick increase in the quantum volume. 
   The results presented correspond to $^{39}$K atoms driven resonantly by an optical lattice potential, created by laser radiation with a wavelength $10.6\;\mu$m, which oscillates with a frequency 5.46~kHz and with Bragg pulses of wavelength 266 nm.
}
\label{fig4}
\end{figure*}

If we want to perform $\sigma_x$ or $\sigma_y$ operations on a selected qubit, we have to wait for the moment when two wave-packets corresponding to the given qubit meet during the evolution along the periodic trajectory. At that moment, we activate the appropriate focused laser beam along with a second broad laser beam, allowing Bragg scattering and thus the transfer of an atom from one wave-packet moving with momentum $p$ to another wave-packet moving with momentum $-p$, or vice versa [Fig.~\ref{fig3}(a)]. Phase control between the laser beams allows choosing whether the transfer corresponds to $\sigma_x$ or $\sigma_y$ operations, the composition of which enables the realization of $\sigma_z$ operations. Very high fidelity of single-qubit operations is achieved if the full transfer of atoms between wave-packets is divided into a few encounters (see Sec.~\ref{concrete} and Fig.~\ref{fig4}). The period of the resonant trajectory can be regarded as a single clock cycle of the quantum processor. This means that a few cycles of the processor clock are needed to perform high-fidelity single-qubit operations.

So far, we have assumed that the atoms do not interact. Interaction arises if we change the internal state of the atoms. The previously described focused laser beams allow for the local activation of interactions between atoms when the wave-packets they occupy meet during the evolution along the periodic trajectory. This enables us to implement a controlled-Z gate between any pair of qubits in the following way.
Let us assume that one atom occupies the wave-packet corresponding to the state $|1\rangle$ of a certain qubit, and another atom occupies the wave-packet corresponding to the state $|1\rangle$ of another qubit. Just before the moment when these two wave-packets meet, we can change the internal state of the atoms by activating the appropriate focused laser beams together with the broad laser beam, causing Raman transitions and altering the internal state of the atoms [Fig.~\ref{fig3}(b)]. Then, during the passage of the wave-packets, the atoms interact. Immediately after the passage, we use laser beams again to restore the atoms to their initial internal states, in which there are no interactions. If the interaction energy is properly chosen, a phase imprint of $e^{i\pi}$ occurs during the interaction. Such a phase imprint only occurs when both qubits are in the $|1\rangle$ states, thus implementing the controlled-Z gate. To achieve high-fidelity controlled-Z gates, the entire gate operation needs to be divided into a few encounters of wave-packets (see Sec.~\ref{concrete} and Fig.~\ref{fig4}), i.e., a few clock cycles of the quantum processor are necessary. All pairs of wave-packets meet each other during a single period, $sT$, of the resonant trajectory. Since the controlled-Z gate implementation does not alter the wave-packet populations of atoms, it can be executed simultaneously even for all possible qubit pairs, i.e., for $s/2(s/2-1)$ pairs, significantly reducing quantum computation time. The ability to perform two-qubit operations between all qubits makes a key parameter of quantum computing, i.e., the quantum volume \cite{Cross2018ValidatingQC}, achieve very high values, especially when the Raman transfer error is reduced [see Sec.~\ref{concrete} and Fig.~\ref{fig4}(d)].

The final stage of the quantum computation is the readout of information from the system’s final state. This can be achieved by measuring the atomic populations of the wave packets — further details are provided in Appendix~\ref{appb}.

We emphasize that even with only a few focused laser beams, we can perform a network of quantum operations that are challenging to achieve in other experimental setups. A single beam focused in one location is sufficient to execute all necessary single-qubit operations, but in this case, the qubits must be associated with wave-packets that meet at the focus point of the beam. The addition of a second beam focused in another location allows the implementation of a controlled-Z gate between the nearest neighboring qubits that can be arranged in a 1D chain. If we introduce yet another focused beam, we gain the capability to perform controlled-Z gates between nearest neighboring qubits that can be arranged in a square network. A fourth beam enables controlled-Z gates between the nearest neighboring qubits arranged in a 3D cube, and so on.

Although our focus is primarily on bosons, the same scheme applies to fermions, which offer advantages for quantum computing due to Fermi statistics and the absence of $s$-wave scattering in a single spin state. Controlled Raman transfers enable the creation of spin mixtures, with scattering confined to selected sites. Because atoms remain in the lowest-energy state except during Raman transfers and controlled-Z gate operations, the approach is not limited by coherence times associated with spontaneous emission.

\subsection{An example of a specific realization}
\label{concrete}

As an example of a concrete realization of the system, let us consider $^{39}$K atoms confined in a 1D box potential and subjected to an oscillating optical lattice potential created by CO$_2$ laser radiation with a wavelength of 10.6~$\mu$m. We assume that the 1D confinement is achieved using a strong harmonic trapping potential with a transverse confinement frequency of the order of $\omega_\perp=2\pi\times 100$~kHz. The oscillation of the optical lattice has an amplitude of $\lambda=30$ (60 recoil energies) and a frequency of $\omega=60$ ($2\pi\times 5.46$~kHz). The {\it natural} tunneling is very weak, requiring more than $10^6$ cycles of the driving force to transfer the atom between neighboring wave-packets. However, Bragg scattering can significantly reduce this time. By employing two laser beams with a wavelength of 266~nm (and a waist of the focused beam of $1.9\;\mu$m) propagating at angles $\pm 46^{\circ}$ with respect to the $z$-axis, atom transfer between any wave-packets can be completed within a few meetings of the wave-packets along the resonant trajectory. Figure~\ref{fig4}(a) illustrates the efficacy of such transfers. Faster transfers necessitate stronger Bragg pulses (i.e., larger $\lambda_{\rm Bragg}$), leading to unwanted coupling with other states of the system. Decreasing the intensity of the Bragg lasers reduces the error, but there is a critical point where the error starts increasing due to the effectiveness of the {\it natural} tunneling.

A single focused laser beam (together with a broad beam) is sufficient to perform single-qubit operations for all qubits if the qubit states are assigned so that the wave-packets corresponding to $|0\rangle_i$ and $|1\rangle_i$ pass each other at different moments of time but at the same location for all $i$, i.e., all qubits. On the other hand, if multiple focused laser beams are available, the appropriate assignment allows for parallel single-qubit operations.

For $^{39}$K atoms, which we use as an example here, the s-wave scattering length can be set to zero if the atoms are initially prepared in the hyperfine state $|1,+1\ra$ and the magnetic field is set to the zero crossing value of 350.5~G, in the vicinity of the Feshbach resonance at 402.5~G \cite{DErrico2007,Giergiel2020}. However, if we transfer the atoms to the state $|1,-1\ra$,  which is 213 MHz above the $|1, +1\ra$ state, using a Raman pulse, the corresponding scattering length becomes $-29$ Bohr radii, and the interaction coefficient (\ref{uij}) does not vanish. By varying the frequencies $\omega_\perp$ of the harmonic trap in the transverse directions (i.e., choosing different values of $g_0$), we can analyze how many meetings of the two wave-packets are needed to imprint a $\pi$ phase in the state in which each of the two atoms occupies one of the wave-packets. Simultaneously, we can determine the fidelity of the process, which is the probability that the atoms remain in their initially occupied wave-packets. The results of this analysis are presented in Figs.~\ref{fig4}(b)-\ref{fig4}(c). They correspond to $\omega_\perp/2\pi$ in the range between 50~kHz and 700~kHz.

For the parameters chosen and for the fidelity, $F_R=0.999933$, achieved in the Raman transfer in Ref.~\cite{Bluvstein2022}, the optimal fidelity of the controlled-Z gate is 0.993 which requires 7 cycles of the resonant trajectory [Figs.~\ref{fig4}(b)-\ref{fig4}(c)]. The fidelity was determined by multiplying the fidelity obtained assuming perfect Raman transfers [black curves in Figs.~\ref{fig4}(b)-\ref{fig4}(c)] by $F_R^{8n}$, where the $8n$ comes from the fact that for each of $n=7$ cycles two interactions take place where two atoms are transferred twice by Raman pulses. Higher fidelity of the gate can be achieved either by improving $F_R$ or by increasing the frequency of the optical lattice oscillation $\omega$. A higher $\omega$ allows for a greater oscillation amplitude $\lambda$, thereby enabling the use of stronger interactions without undesired couplings to other states of the system. Stronger interactions facilitate imprinting the $\pi$ phase onto states with high fidelity in fewer cycles of the resonant trajectory, thereby reducing the impact of imperfect Raman transfers.

A significant advantage of the proposed quantum computer is the ability to perform two-qubit gates between any pair of qubits. This is quantitatively illustrated by the quantum volume presented in Fig.~\ref{fig4}(d). In the example we consider here and for the Raman transfer fidelity achieved in Ref.~\cite{Bluvstein2022}, $F_R=0.999933$, and for 49 cycle single-qubit gates [cf. Fig.~\ref{fig4}(a)], a system of 7 qubits is capable of executing universal quantum operations. In this case, achieving the $14:1$ resonance is necessary, and for $^{39}$K atoms, any combination of the $\sigma_x$ and $\sigma_y$ operations takes 126~ms, while the controlled-Z gate operation takes 36~ms. We note that during these periods all single-qubit operations and controlled-Z gate operations, even for all possible qubit pairs, can be executed simultaneously, significantly reducing quantum computation time.

\section{Summary and conclussions}
\label{sandc}

Periodically perturbed systems can give rise to crystalline structures in time, which, in the time domain, exhibit phenomena similar to those known in traditional spatial crystals \cite{SachaTC2020}. In this study, we demonstrate that time crystalline structures possess significant potential for practical applications. In conventional spatial structures, various components of the system are spatially separated, and selective communication and interaction between distant parts of the system are preceded by their spatial transport. In time crystalline structures, the transport problem is automatically resolved because wave-packets associated with sites of time lattices evolve periodically in time, naturally encountering each other at different time instances. This opens up the possibility of creating any temporal printed circuit board capable of implementing a wide range of quantum devices for bosons or fermions. Similar to electronics, this gives rise to the field of time-tronics. What is a formidable challenge in spatial structures becomes trivial in temporal structures. As a concrete example, we describe the implementation of a universal quantum computer, in which the key problem of qubit transport \cite{Cirac2000,Bluvstein2022} for realizing two-qubit gates between any pair of qubits is automatically addressed.

In this article, we present a detailed realization of a temporal printed circuit board and a quantum computer in a system of atoms moving periodically in a box potential. An advantage of such a potential is that all gate operations are performed in the same manner, regardless of where in space the atoms meet. However, a temporal printed circuit board and a quantum computer can also be implemented when atoms move in different potentials. In such cases, the parameters of gate operations should be appropriately chosen depending on where in space the atoms meet. This opens up the possibility of developing time-tronics in a broad class of ultra-cold atomic gas laboratories.

\section*{Acknowledgments}
We acknowledge Michal Grabowski for preparing the graphics.
This research was funded by the National Science Centre, Poland, Project No. 2021/42/A/ST2/00017 and the Australian Research Council Discovery Project Grants DP190100815 and DP240101590. 

\appendix
\section{Counter-diabatic driving to the initial state}
\label{appa}

This appendix details the procedure for preparing the initial state in a temporal printed circuit board. The method is broadly applicable, including to quantum computing implementations and other experiments based on the temporal circuit board.

\begin{figure}[t]
\begin{center}
\includegraphics[width=0.99\columnwidth]{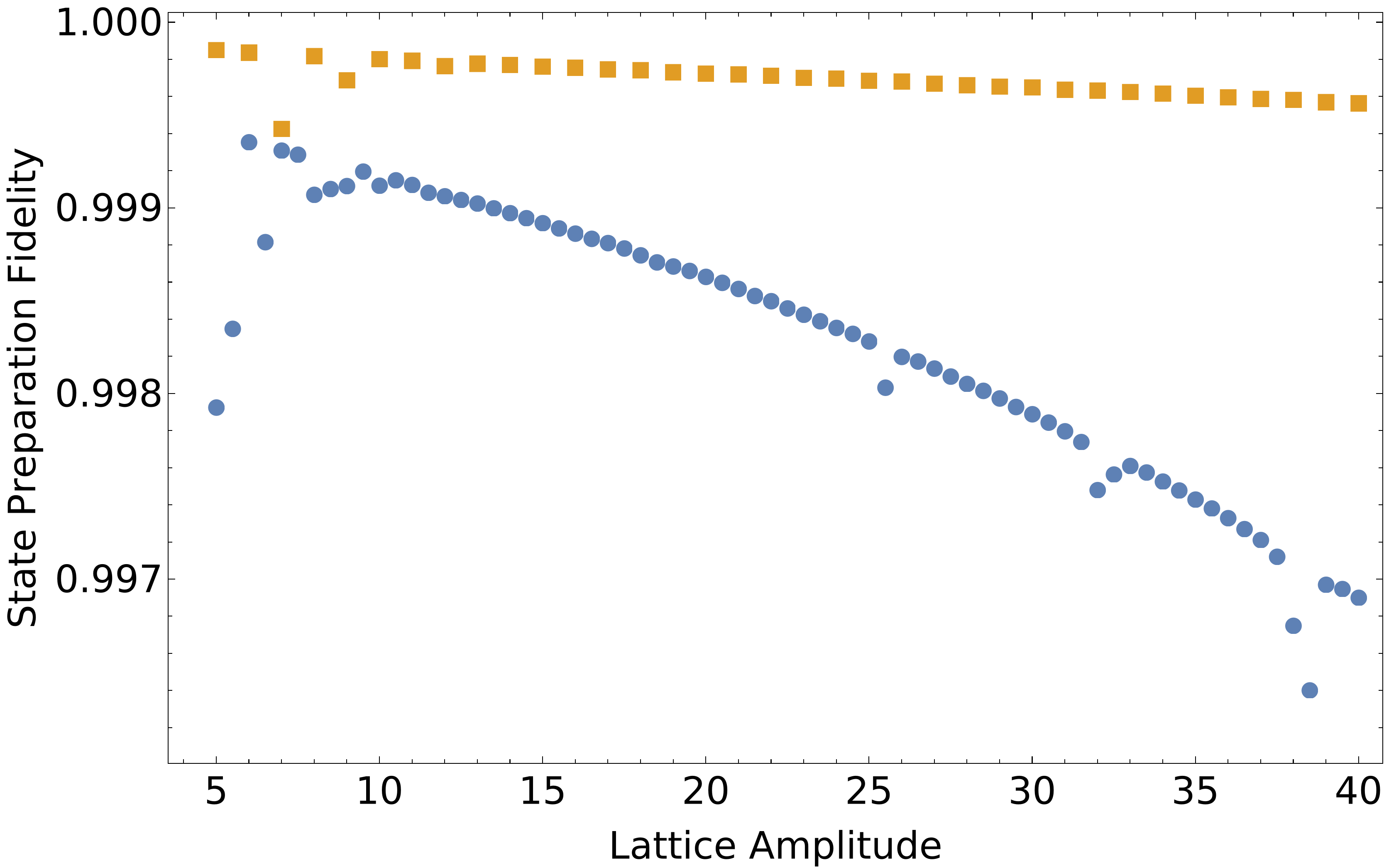}    
\end{center}
\caption{{\bf Fidelity of the initial state preparation.}
Initially, the atom is prepared in the ground state of a static optical lattice potential. It is then accelerated to the resonant momentum $p_{\rm res} = \omega/2$ using counter-diabatic driving. The figure shows the fidelity of the desired final state as a function of the amplitude of the optical lattice oscillations. Blue dots correspond to the resonant momentum $p_{\rm res} = 30$, while orange squares correspond to $p_{\rm res} = 60$.
}
\label{SUPLattice}
\end{figure}

In Sec.~\ref{ideaqc}, we have outlined the procedure for preparing the initial state in the quantum computer. This process consists of several steps. Initially,
$N=s/2$ atoms are prepared in a Mott insulator state within a time-independent potential composed of a box potential and 
$s/2$ potential wells created by a standing electromagnetic wave --- each well contains exactly one localized atom. Next, we can turn off interactions between atoms by tuning an external magnetic field near a Feshbach resonance such that the scattering length becomes zero. We then set the atoms in motion with the momentum value, $p_{\rm res}$, that will satisfy the resonance condition once the oscillation of the optical lattice potential is activated. In this section, we describe how to initiate atomic motion with the correct momentum quickly and with high fidelity. The method we present is an example of counter-diabatic driving \cite{Guery2019}, which quickly enables setting the atoms in motion both in the presence and absence of interactions. For simplicity, we will assume non-interacting atoms.

Let us consider a single atom prepared in the lowest energy state, $\psi(z)e^{-iEt}$, in one of the potential wells of the static optical lattice, assuming negligible tunneling between wells. The atom is described by the Hamiltonian 
\be
H_0=\frac{p^2}{2}+\lambda \cos^2(z).
\label{H0SM}
\ee
Our goal is to set the optical lattice potential in motion in such a way that the atom follows the motion of the potential well in which it is localized with the momentum $p_{\rm res}$ that matches the desired resonant value. The presence of a moving optical lattice potential ensures that the atom remains in the ground state of the well in the reference frame co-moving with the well. 
Setting the optical lattice in motion along a trajectory $z_0(t)$ can be achieved using a simple unitary transformation
\begin{eqnarray}
    U(t)&=&
    \exp\left[izz_0'(t)\right]
    \exp[-ipz_0(t)],
    \label{UtSM}
\end{eqnarray}
where $z_0'=dz_0/dt$, which leads to the Schr\"odinger equation with a new Hamiltonian
\be
H_0=\frac{p^2}{2}+\lambda \cos^2[z-z_0(t)]-zz_0''(t),
\label{H0pSM}
\ee
where $z_0''=d^2z_0/dt^2$ and an unimportant purely time dependent term has been omitted.

Initially, we have the original Hamiltonian (\ref{H0SM}), and at the final moment, we want to ensure that $z_0(t)=p_{\rm res}t$ and consequently $z_0''(t)=0$. The unitary transformation (\ref{UtSM}) provides a prescription for how we should change the Hamiltonian (\ref{H0pSM}) over time to achieve the desired final result. This result can be obtained by choosing different trajectories $z_0(t)$ and transitioning from the initial Hamiltonian to the final one at any speed. The unitary transformation (\ref{UtSM}) guarantees that, regardless of how quickly the transition is made, the transformed solution will always satisfy the Schr\"odinger equation and, at the final moment when $z_0(t)=p_{\rm res}t$, will correspond to a wave-packet moving with the desired momentum
\begin{equation}
    U(t)\psi(z)e^{-iEt}=\psi(z-p_{\rm res}t)e^{izp_{\rm res}}e^{-iEt}.
\end{equation}
If we want to quickly execute the procedure for preparing a moving wave-packet, we need to use a stronger linear potential with a gradient equal to the acceleration $z_0''(t)$. Such a potential can be created using an inhomogeneous magnetic field. 
For example, for the parameters used in Sec.~\ref{concrete}, the preparation of the initial state for quantum computing requires that the initially static optical lattice potential is accelerated to the velocity 29~mm/s which can be achieved by introducing an appropriate frequency difference between the the counter-propergating 10.6 micron laser beams. If the preparation process is realized during 5~ms, the gradient of the linear potential needed to perform the counter-diabatic driving can be created using a magnetic field gradient of about 10 G/cm. Regardless of how quickly the excitation procedure is performed, it will always remain equally precise.

At the end of the process, we replace the moving optical lattice with an optical lattice oscillating at the resonant frequency with the amplitude $4\lambda$, thereby obtaining the desired wave-packet evolving along the resonant trajectory. Within the secular approximation (or rotating wave approximation) \cite{Buchleitner2002,SachaTC2020}, in the frame co-moving resonantly with the atom, the optical lattice appears exactly as it does in the Hamiltonian (\ref{H0SM}). In a more precise description, there are also rapidly oscillating terms, which we neglect under the secular approximation. Their presence causes the moving wave-packet obtained via the counter-diabatic driving described above to be slightly different from the exact resonant state. For the parameters chosen in this article, the fidelity is 99.8\%, and it can be made arbitrarily close to 100\% by selecting higher resonant momenta, i.e., higher frequencies of the optical lattice oscillations. In Fig.~\ref{SUPLattice} we present the fidelity versus the amplitude of the lattice oscillations for two different oscillation frequencies $\omega$.

\begin{figure}[h]
\includegraphics[width=0.99\columnwidth]{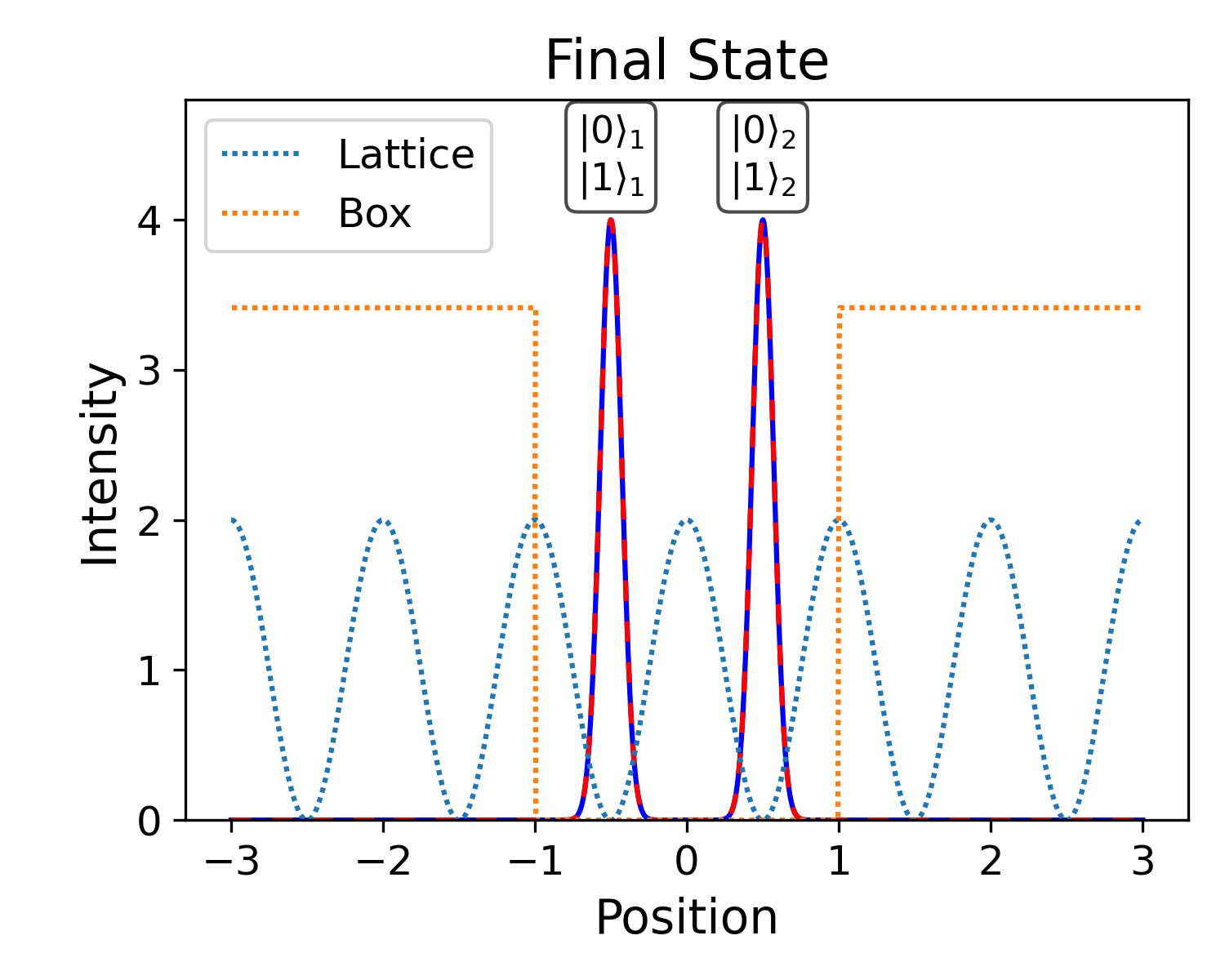}
\hfill
\includegraphics[width=0.99\columnwidth]{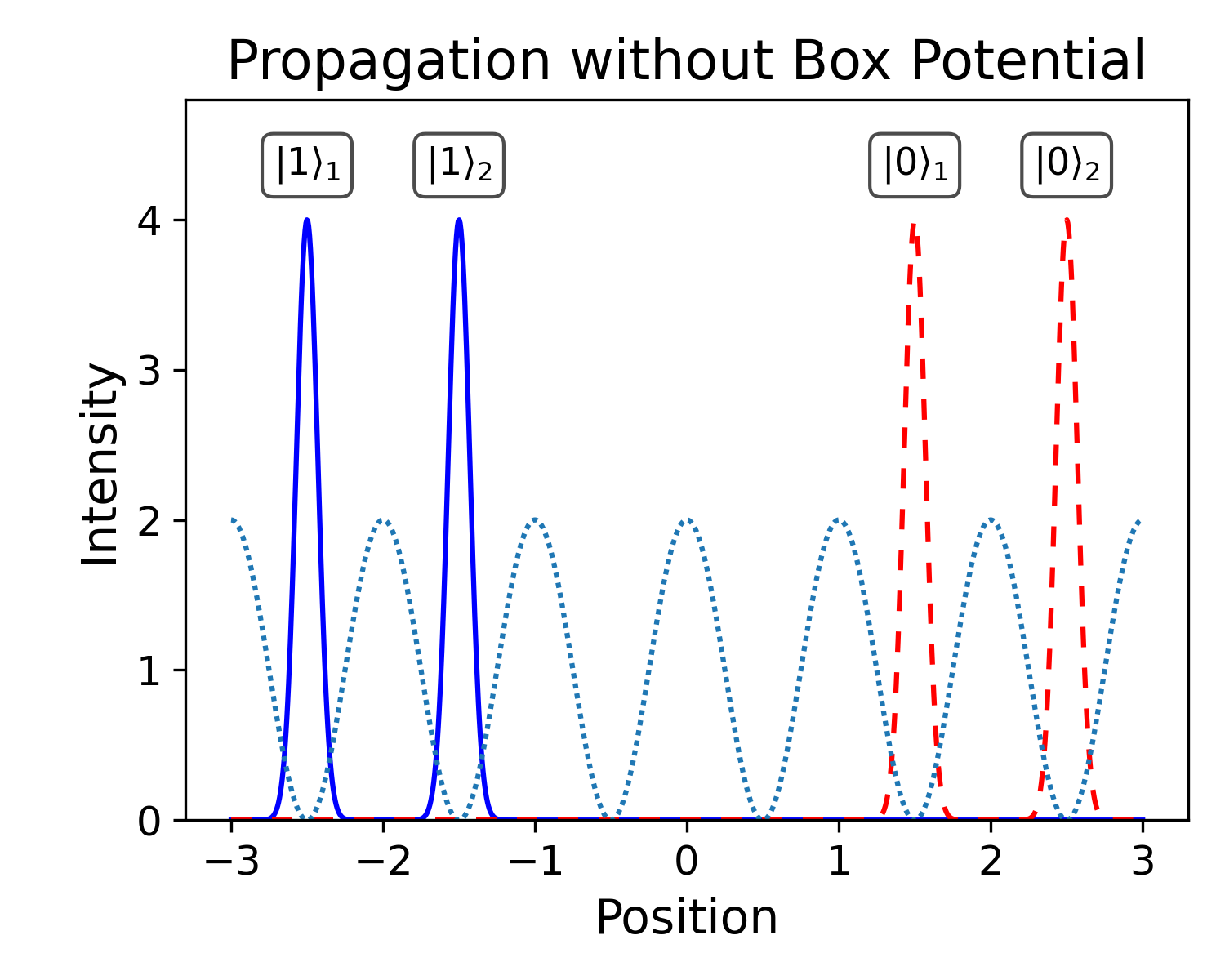}
\caption{{\bf State readout protocol:}
    {\bf (Top)}: Schematic of the final states for a two-qubit quantum computation. Assuming we have two qubits, i.e., each of the two atoms has a wave-packet $|0\rangle$ or $|1\rangle$ at its disposal. Let us assume that the final moment of quantum computation has been chosen so that the wave-packets $|0\rangle$ move to the right and the wave-packets $|1\rangle$ move to the left. Blue dotted line indicates the oscillating optical lattice potential while orange dotted line represents the box potential. {\bf (Bottom)}: After turning off the box potential, wave-packets $|0\rangle$ and $|1\rangle$ evolving in the presence of the oscillating lattice potential separate from each other. The smallest distance between the wave-packets is 5.3~$\mu$m and their population by atoms can be imaged with high fidelity using standard atom fluorescence imaging.
}
\label{figS1}
\end{figure}

\section{Detection of the final state in quantum computations}
\label{appb}

A crucial step in quantum computation is a reliable readout of the final quantum state. The process is illustrated in Fig.~\ref{figS1}, which shows how the qubit states $|0\rangle_i$ can be spatially separated from the $|1\rangle_i$ states and then their populations by atoms can be observed using standard atom imaging techniques. 

For simplicity, let us assume that as the detection moment we have chosen the moment of time when the wave-packets corresponding to the states $|0\rangle_i$ are moving to the right while the wave-packets corresponding to the states $|1\rangle_i$ are moving to the left.

To initiate spatial separation, the confining box potential is removed while the oscillating lattice potential remains active. As a result, the wave-packets corresponding to different logical states begin to move apart in space.

After the left- and right-moving wave-packets are separated their population by atoms can be measured. Because the smallest distance between the wave-packets is on the order of $5.3~\mu\text{m}$ for the parameters chosen in this paper, individual atoms can be imaged with high fidelity using standard atom fluorescence imaging. The presence of fluorescence indicates which atoms are in the $|0\rangle_i$ state and which are in $|1\rangle_i$ state which is the information we need at the end of the quantum computation.

A similar process, where the barriers are temporally turned-off can be used to facilitate mid-circuit measurements. 
\vfill


%

\end{document}